\documentclass[a4paper,fleqn,usenatbib]{mnras}
\usepackage{ae,aecompl}

\usepackage{natbib}
\usepackage{graphicx}
\usepackage{bm}
\usepackage{amssymb}
\usepackage{amsmath}
\usepackage{cancel}
\usepackage{xcolor}

\usepackage{hyperref}

\usepackage{epstopdf}
\usepackage{epsfig}

\newcommand{\avg}[1]{\ensuremath{\langle #1 \rangle}}
\newcommand{\bma}{\begin{math}}
\newcommand{\ema}{\end{math}}
\newcommand{\beq}{\begin{equation}}
\newcommand{\eeq}{\end{equation}}
\newcommand{\beqa}{\begin{eqnarray}}
\newcommand{\eeqa}{\end{eqnarray}}
\newcommand{\bc}{\begin{center}}
\newcommand{\ec}{\end{center}} 
\newcommand{\bit}{\begin{itemize}}
\newcommand{\eit}{\end{itemize}}

\newcommand{\y}{$y$~}

\font\BFd=cmmib10
\font\BFt=cmmib10
\font\BFs=cmmib10 scaled 700
\font\BFss=cmmib10 scaled 500

\def\bbox#1{%
\relax\ifmmode
\mathchoice
{{\hbox{\BFd #1}}}
{{\hbox{\BFt #1}}}
{{\hbox{\BFs #1}}}
{{\hbox{\BFss #1}}}
\else \mbox{#1} \fi }

\title{A Measurement of the Galaxy Group-Thermal Sunyaev-Zel'dovich Effect Cross-Correlation Function}
\author[Vikram, Lidz \& Jain]{Vinu Vikram$^{1,2}$\thanks{E-mail: vvinuv@gmail.com}, Adam Lidz$^{2}$, Bhuvnesh Jain$^{2}$ \\
$^{1}$Argonne National Laboratory, 9700 South Cass Avenue, Lemont, IL 60439, USA\\
$^{2}$Department of Physics and Astronomy, University of Pennsylvania, Philadelphia, PA 19104, USA\\}
\date{To be submitted to \mnras.  Received ; in original form } 

\begin{document}
\maketitle

\begin{abstract}
Stacking cosmic microwave background (CMB) maps around known galaxy clusters and groups provides a powerful probe of 
the distribution of hot gas in these systems via the Sunyaev-Zel'dovich (SZ) effect. A stacking analysis allows
one to detect the average SZ signal around low mass halos, and to extend measurements out to large scales, which are too
faint to detect individually in the SZ or in X-ray emission. In addition, cross correlations between SZ maps
and other tracers of large-scale structure (with known redshifts) can be used to extract the redshift-dependence of the SZ background.
Motivated by these exciting prospects, we measure the two-point cross-correlation function between a catalog of
$\sim 380,000$ galaxy groups (with redshifts spanning $z=0.01-0.2$) from the Sloan Digital Sky Survey (SDSS) and Compton-\y parameter
maps constructed by the Planck collaboration.
We find statistically significant correlations between the group catalog and  
Compton-\y maps in each of six separate mass bins, with estimated halo masses in the range 
$10^{11.5-15.5} M_\odot/h$. We compare 
these measurements with halo models of the SZ signal, which describe the stacked measurement in terms of one-halo and two-halo terms. 
The one-halo term quantifies
the average pressure profile around the groups in a mass bin, while the two-halo term describes the contribution of correlated
neighboring halos. For the more massive groups we find clear evidence for the one- and two-halo regimes, while groups with mass below $10^{13} M_\odot/h$ are dominated by the two-halo term given the resolution of Planck data. Higher angular resolution CMB data
is required to probe the hot gas content of low mass halos using our method. 
We use the signal in the two-halo regime to determine the bias-weighted electron pressure of the universe:  $\avg{b P_e} = 1.50 \pm 0.226 \times 10^{-7}$ keV cm$^{-3}$ (1-$\sigma$) at $z\approx 0.15$. 
\end{abstract}


\section{Introduction} \label{sec:intro}

The Sunyaev-Zel'dovich (SZ) effect \citep{Sunyaev72} refers to the spectral distortion generated as cosmic microwave background (CMB)
photons inverse Compton scatter off of hot free electrons along the line of sight.
This effect has traditionally been used
to study the abundance and hot gas properties of galaxy clusters, the most massive bound objects in the universe (e.g. \citealt{Carlstrom:2002na,Ade:2015fva} and references therein).
Recent measurements, however, have detected an SZ signal around
lower mass halos by stacking CMB maps around known galaxy groups and bright galaxies, extending
the cluster-scale measurements to lower mass systems that are too faint to be
detected individually in the SZ. First, \citet{Hand:2011ye} stacked Atacama Cosmlogy Telescope (ACT) data
around Luminous Red Galaxy (LRG) samples from the Sloan Digital Sky Survey (SDSS) and detected the SZ signal. More recently, the Planck Collaboration 
detected the SZ signal around systems with stellar masses as small as $M_\star \sim 2 \times 10^{11} M_\odot$ \citep{Ade:2012nia}. The Planck
analysis has recently been independently revisited and refined by \citet{Greco15}, who reach mostly similar conclusions.

These measurements provide empirical constraints on the abundance and spatial distribution of the hot gas within dark matter halos,
as a function of the host halo mass, and thereby test a basic prediction of galaxy formation models. Constraints on
the ``hot phase'' can then be combined with measurements of the cold gas content and stellar mass-halo mass relations to build up a complete census of the
baryonic content of halos of varying mass. Future measurements, with halo mass
tracers at various redshifts, should further determine the redshift dependence of the hot gas content of dark matter halos.
A stacking analysis might also help reveal gas in the warm-hot intergalactic medium (WHIM) from shocked, filamentary
gas within and outside halos; this phase remains elusive even though it is thought to harbor roughly
half of the baryons in the universe at the present day \citep{Cen:1998hc,Fukugita:2004ee}.

Intriguingly, the  Planck  measurements are consistent with the
cluster/group thermal energy scaling with halo mass as a pure $\propto M^{5/3}$ power-law, across roughly
three decades in halo mass \citep{Ade:2012nia}. This is the scaling expected
if the baryonic thermal energy arises solely from shock heating as the cluster baryons fall in and thermalize in the potential
wells set up by the dark matter. On the other hand, it has long been known that groups and clusters do not obey this nearly self-similar
scaling in detail; some combination of radiative cooling and feedback from AGN and supernovae is likely responsible (e.g., \citealt{Kaiser86,Kaiser91}, and reviewed by
\citealt{Voit:2004ah}). In particular, X-ray observations show that groups are less-luminous in X-rays -- given their temperatures -- than expected from a self-similar
scaling, and have lower gas fractions and higher entropies than in the self-similar case (e.g. \citealt{Osmond:2004rd,Sun:2008eh}). In current galaxy formation models, this is explained mostly
by AGN feedback. The AGN feedback acts to push gas out of the inner regions of these systems -- and in some cases ejects gas from the halo entirely -- lowering the X-ray
luminosity, the gas fraction, and raising the entropy; it also reduces the model star formation rates and helps with the ``over-cooling'' problem (e.g. \citealt{Puchwein:2008dh,Brun:2013yva}).
These effects are particularly prominent in low mass groups, which show especially strong departures from self-similarity.

Recent work has, in part, addressed this empirically by stacking X-ray data from the ROSAT survey around the same catalog of clusters/groups/central galaxies \citep{Anderson:2014jxa} as used in the Planck SZ analysis. These authors find significant evidence for non-gravitational heating in the X-ray data. They suggest that this may be reconciled with the Planck results provided that small mass halos
do in fact retain close to the cosmic mean baryon fraction in hot gas, and provided the gas in low-mass systems is less centrally-concentrated so that it emits weakly in X-rays.
In any case, the recent results clearly invite further investigation, both in terms of comparing the measurements with theoretical models \citep{Brun:2015ima,vandeVoort:2016qgv}, and in testing and developing associated data
analysis procedures \citep{Greco15}.

Toward this end, we attempt a similar measurement to that of the Planck collaboration, but we employ a different group catalog and a rather different analysis technique.
The Planck team used a matched-filter approach to estimate a single number (as a function of stellar/halo mass):
$Y_{500}$, the Compton \y-parameter integrated out to $\theta_{500}$, the angle spanned by the radius ($R_{500}$) at which each halo
encloses an average density of $500$ times the critical density. One issue with this approach relates to the angular resolution of Planck, which implies that the measurement is sensitive only 
to the SZ flux on larger angular scales. The Planck collaboration then assumes a simple ``universal pressure profile'' \citep{Arnaud10} to extrapolate their results inward to $\theta_{500}$; this extrapolation may be unreliable
if AGN feedback impacts the pressure profile significantly \citep{Brun:2015ima}.

In this work, rather than estimating only $Y_{500}$, we  measure the full two-point cross-correlation,
i.e., the average y-profile around galaxy groups
as a function of radial separation, in several bins of host halo mass. Thus we make use of the
full scale-dependence of the stacked signal, which can help to separate the pressure profile around the groups of interest and
the two-halo contribution from correlated neighboring systems. 
The two-halo term is itself interesting, as it depends primarily on the average bias-weighted pressure of the
universe; this includes contributions to the thermal energy from {\em all systems} (as opposed to only those above
some observational flux limit), although at the redshifts considered here this quantity is dominated by massive halos,
as we shall see. Our work overlaps here with the recent studies by \citet{waerbeke14} and \citet{hill14}. These works considered the cross-correlation
of SZ maps with CFHTLenS data and CMB lensing maps, respectively. The broad lensing kernel means that these
previous measurements are less-localized in redshift than in our analysis which uses a catalog of groups with
known redshifts (from $z=0.01-0.2$ although peaked around $z \sim 0.1-0.2$ for most halo mass bins, see Fig. \ref{fig:zdist-group}).   
In addition, these
studies -- especially \citet{hill14} -- are sensitive to the average bias-weighted pressure at higher
redshift than our measurement.

The outline of this paper is as follows. In \S \ref{sec:data}, we describe the SDSS group catalog and Planck 
data used in our analysis. 
The correlation function measurements are presented in \S \ref{sec:measure}, 
where we also briefly describe the halo models used to help interpret the measurements. 
In \S \ref{sec:systematics}, we test our measurements for systematic contamination, including that from cosmic infrared background (CIB) radiation that
leaks into the Planck \y maps.
In \S \ref{sec:comparison}, we discuss previous related measurements, especially the Planck analysis of \citet{Ade:2012nia}. In \S \ref{sec:future} we comment on the prospects for improving these 
measurements in the future. 
We conclude in \S \ref{sec:conclusion}.
Throughout, we assume a \(\Lambda\)CDM model with
\(n_{s} = 1\), \(\sigma_{8} = 0.8\), \(\Omega_{m} = 0.27\),
\(\Omega_{\Lambda}=0.73\), \(\Omega_{b}=0.044,\) and
\(h=0.7\), broadly consistent with recent parameter determinations \citep{Ade:2013zuv}.

\section{Data Sets and Measurement Technique}
\label{sec:data}

\subsection{SDSS Group Catalog}
\label{sec:group}

Our study uses the galaxy group catalog of \citet{yang07} , constructed from the SDSS DR4 spectroscopic galaxy survey. This catalog consists of groups of galaxies that are identified as likely to reside in the same dark matter halo.  It provides an estimate of the center of each group, along with two estimates of the underlying halo mass: one determined from the luminosity of the member galaxies, and one
based on stellar mass. The group finder has been tested extensively with mock galaxy catalogs, and successfully finds isolated galaxies in smaller mass halos as well as richer groups and clusters. In general, we refer to all systems identified by the catalog loosely as ``groups'', although some of these are really isolated galaxies, while others are actually full-fledged clusters. For our purposes, the broad dynamic range in halo mass covered by the catalog is
valuable because it allows us to study the SZ effect across this full range in halo mass. In our analysis, we adopt the halo mass estimate based on the luminosity of the member galaxies throughout (as opposed to the stellar-mass based estimate). \citet{yang07} show 
that the two estimates agree well (their Figure 10) and find that the scatter between the two estimates is small compared to the overall spread in the member luminosity-halo mass relation. Therefore
our results should be insensitive to this choice.

\begin{figure}
\centering
\includegraphics[scale=0.5]{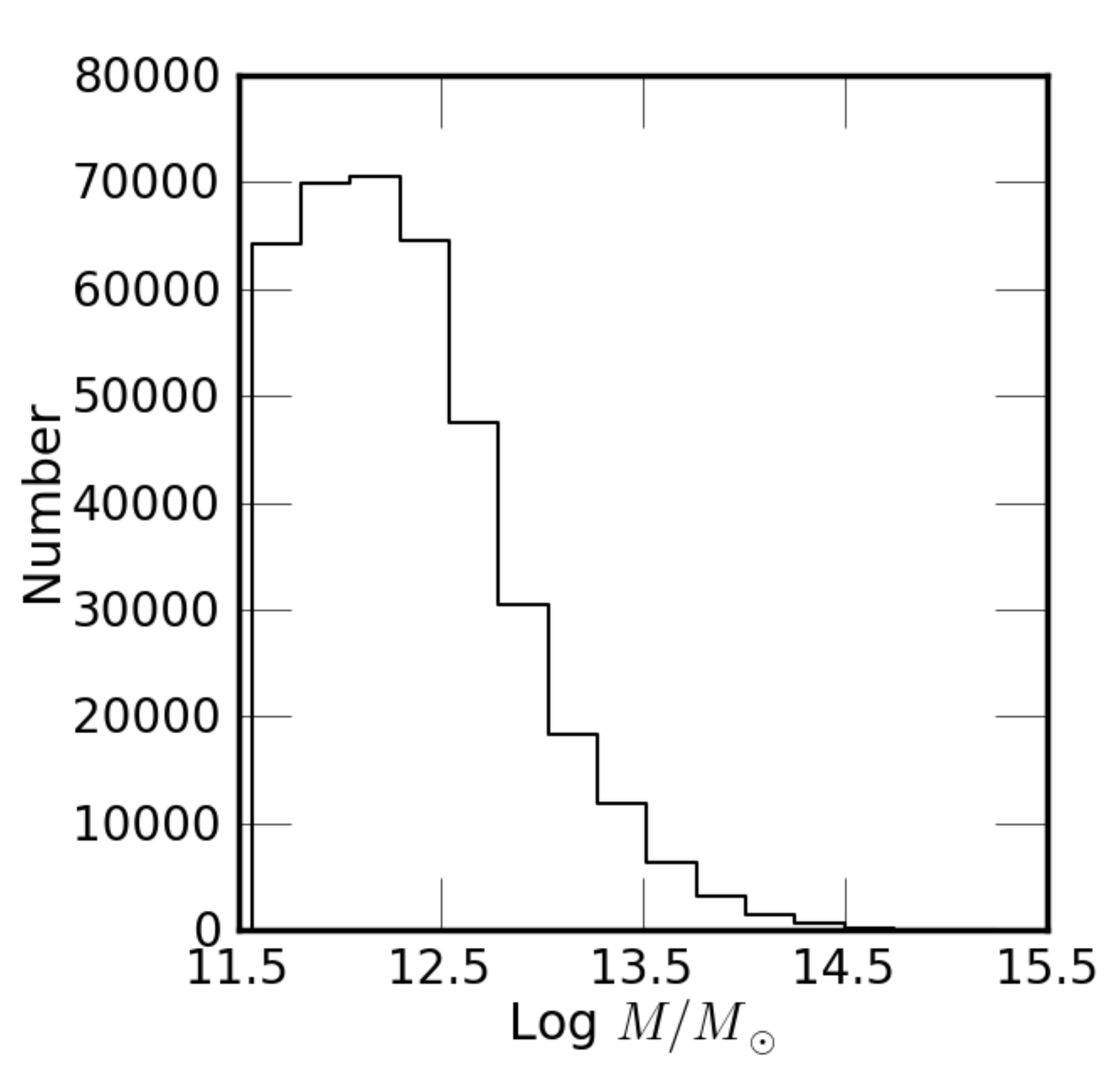}
\caption{Mass distribution of groups in the \citet{yang07} catalog. The $x$-axis is the halo mass estimate from \citet{yang07} based on the luminosity of the member galaxies, while
the $y$-axis shows the number of groups in each mass interval across the full redshift range of the catalog.}
\label{fig:mdist-group}
\end{figure}

\begin{figure}
\centering
\includegraphics[scale=0.5]{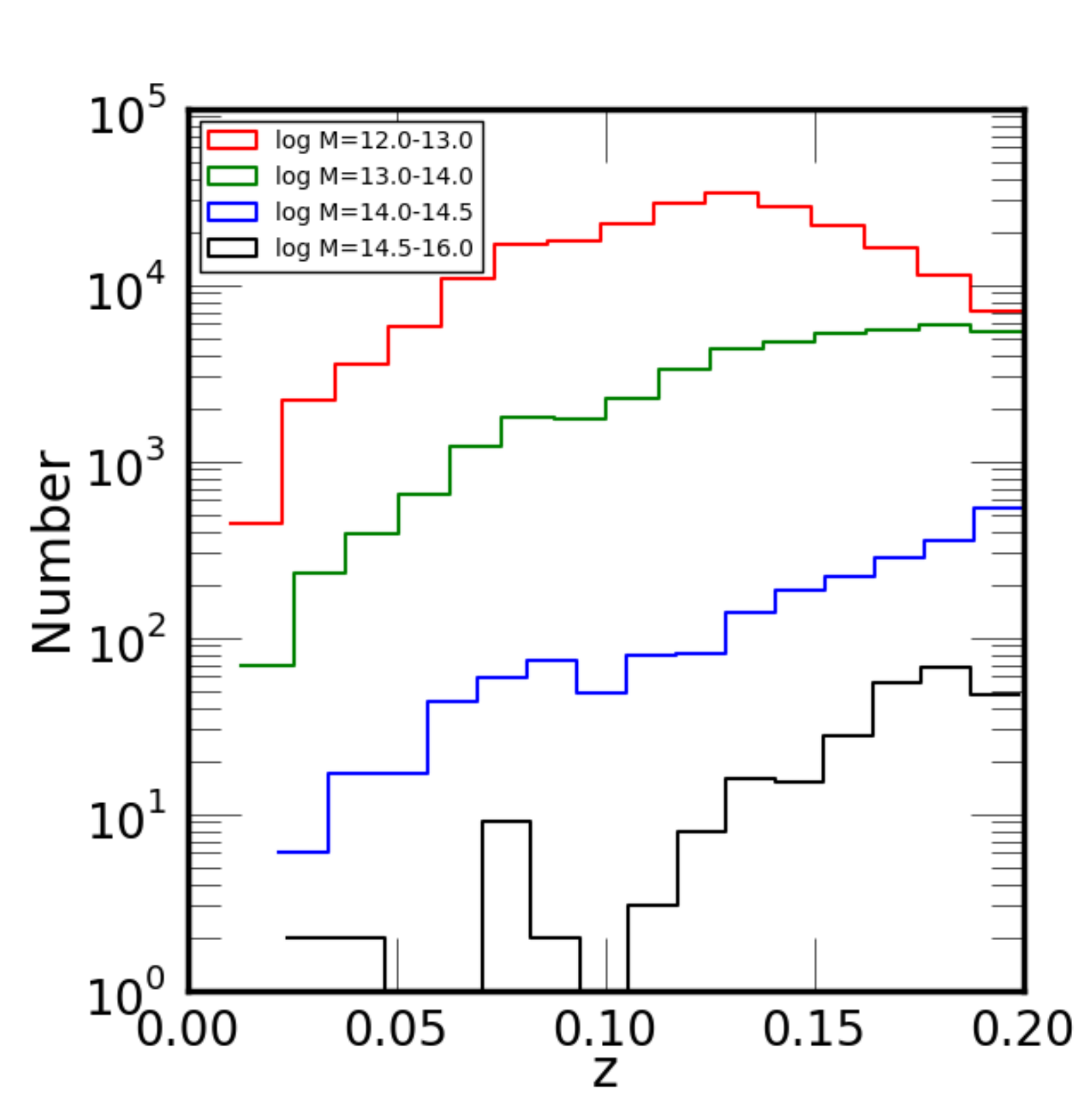}
\caption{Redshift distribution of  groups and clusters in different mass bins from the \citet{yang07} catalog. The histograms show the redshift distribution in each of the six mass bins used in our analysis.}
\label{fig:zdist-group}
\end{figure}

The catalog includes 389,279 groups with halo masses ranging between $10^{11.5} - 10^{15.5} h^{-1} M_\odot$ and redshifts
between $z=0.01-0.2$. For most of our analysis, we divide these systems into six separate halo mass bins, with $\log_{10}(M/(h^{-1} M_\odot)) = (11.5-12.0); (12.0-13.0); (13.0-13.5); (13.5-14.0); (14.0-14.5); (\geq 14.5)$.
The number of objects in each mass bin are given in Table \ref{tab:groups-basic}, while Figs. \ref{fig:mdist-group} \& \ref{fig:zdist-group} show the mass and redshift distributions, respectively, of the groups
in our analysis.

\begin{table}
\begin{center}
 \begin{tabular}{cc}
Log $M/(h^{-1} M_\odot)$ & No of objects \\
\hline
$<$ 12.0 & 118036 \\
12.0-13.0 & 225959 \\
13.0-13.5 & 32606 \\
13.5-14.0 & 10267 \\
14.0-14.5 & 2152\\
$>$ 14.5 & 259\\
\end{tabular}
\end{center}
\caption{Number of groups in each halo mass bin.}
\label{tab:groups-basic}
\end{table}

\subsection{Planck Compton-\y Maps}
\label{sec:ymap}

In order to study the SZ effect around the SDSS groups, we use the publicly available Compton-\y parameter maps produced by the Planck collaboration \citep{Aghanim:2015eva}. These maps provide us with estimates of the Compton-$y$ parameter across almost the entire sky. The Compton-\y parameter, $y$, is defined by the spectral distortion the SZ effect generates on the CMB:
\begin{equation}
\frac{\Delta T}{T_{CMB}} = y g(x),
\end{equation}
where $\Delta T = T-T_{CMB}$, $T$ is the observed CMB temperature in a given frequency, $T_{CMB}=2.726$ K, $g(x)= x ~\mathrm{coth}(x/2) - 4$, and $x=h\nu/k_B T_{CMB}$.\footnote{Note that the Planck collaboration removes the overall mean \y from the map, but only the excess \y -- above the mean -- around groups is of interest here.}  
We use \y maps derived by the Planck collaboration based on
two different SZ-reconstruction techniques, a MILCA (Modified Internal Linear Combination Algorithm) \citet{Hurier13} map, and a NILC (Needlet Internal Linear Combination) \citet{Remazeilles11} map. Each reconstruction uses the known frequency dependence of the thermal SZ effect to separate it from the primary CMB and foreground contaminants. More specifically, linear combinations of 
the multi-frequency Planck maps are taken to minimize the variance of the reconstructed map, while providing unit response to the SZ signal and zero contribution from the primary CMB. In both the MILCA and NILC algorithms, the weights in the linear combinations vary with spatial scale and are computed separately for different regions on the sky. In this way, both spatial and spectral information are used in separating the SZ effect from other contributions to the Planck data, with MILCA and NILC differing in the details of how the weights are constructed \citep{Aghanim:2015eva}. 
The Planck \y maps are derived from linear combinations of the Planck High-Frequency Instrument (HFI) temperature maps at 100 GHz, 143 GHz, 217 GHz, 353 GHz, 545 GHz, and 857 GHz \citep{Aghanim:2015eva}, with each input map first smoothed to a common resolution of FWHM=10 arcmin.\footnote{The NILC map also uses large angular scale data from the Low-Frequency Instrument (LFI) at
$\ell < 300$.}
In this work, we mostly use the MILCA \y-map but  show some results from the NILC map in \S \ref{sec:systematics}. In that section, we show tests suggesting that NILC is less reliable for our purposes. 
In order to further mitigate foreground contamination from our own Milky Way galaxy and from bright point sources, we mask 60\% of the sky based on the Planck galactic mask and the union of
the HFI and LFI point source masks. The remaining sky coverage is then used in estimating the two-point correlation function.

\subsection{Estimating the Two-Point Correlation Function}
\label{sec:measurement}

In order to estimate the \y-group correlation function, we first divide each  group mass-bin into a set of narrow redshift slices of width $\delta z = 0.01$. The \y-group correlation function is estimated
from each redshift slice separately; these results are then combined by weighting each slice by the probability that a group -- from the mass bin of interest -- 
lies in the slice (see Fig.~\ref{fig:zdist-group}). The measurements in each slice are performed using the TreeCorr code of \citet{Jarvis04}.
The individual estimates are combined in this way because the measurements from each individual redshift slice are too noisy to be useful.
Throughout we measure the \y-group correlation as a function of the co-moving separation $r$ from the center of a group. To do this,
we first determine the co-moving distance
from the center of each group to the center of each Planck pixel using the redshift of the group and the co-moving distance to that redshift. 

Our correlation function estimator may be expressed symbolically as:
\begin{equation}
\xi_{y,g}(r) = \sum_i w_i(z_i) \left[\langle D_g(z_i) y(r) \rangle - \langle R(z_i) y(r) \rangle \right].
\label{eq:yest}
\end{equation} 
The sum here runs over the narrow redshift slices: slice $i$ is centered around redshift $z_i$, and $w_i$ gives the probability that a group in the mass bin of interest lies in slice $i$. 
The estimator determines the average $y$, in an annulus of radius $r$, around groups from the SDSS catalog at each redshift -- denoted here by $D_g(z_i)$ -- in excess of that around random points at the
same separation, $R(z_i)$. The
first term in Eq.~\ref{eq:yest} hence denotes the average value of $y$ at a distance $r$ from groups in the catalog, while the second-term gives the average $y$ around points from a random catalog at
the same separation.
We suppress
the dependence on mass-bin here for brevity, but this estimator is applied separately to each of our group mass bins. The averages $\langle D_g(z_i) y(r) \rangle$ and $\langle R(z_i) y(r) \rangle$ are
computed including all groups or randoms in the redshift slice and mass bin with equal weight, and incorporate all Planck pixels in the annular bin with equal weight. In some cases, only a portion
of an annular bin around a group or random point will be outside of the Planck mask, and this is accounted for in computing the average.  

In order to estimate the correlation function robustly, an accurate random catalog
is required. This must incorporate mass and redshift-dependent spatial variations in the selection function of the SDSS groups. Here we make use of the random catalog from \citet{yang07} which
fully accounts for variations in the completeness of the group catalog and its mask. The ratio of the number of random points in the catalog to actual groups depends on the mass bin, but in total -- i.e., across all mass bins --  the number of random points is just a little under twice the number of SDSS groups.

\subsection{Error Bar Estimation}

In order to estimate the error bars on the measured SZ-group correlation function, $\xi_{y,g}(r)$, we use the jackknife resampling technique (e.g. \citealt{norberg09}). Specifically, we divide the sky into $N$
sub-regions, each of size $58$ sq. degrees. We then make $N$ separate estimates of the two-point function: in each estimate, we first omit a different one of the sub-regions from the analysis and then measure the two-point
function from the remaining sample. Let us call the estimate of the two-point function in the $i$th radial bin, after omitting the $k$th sub-region from the analysis, $\xi_i^k$, while let $\bar{\xi_i}$ denote the two-point function estimated from the full data sample, i.e., after averaging over all sub-regions. The jackknife estimate of the co-variance between the
two point function estimates in the $i$th and $j$th radial bin is then given by:

\begin{equation}
cov(\xi_i, \xi_j) = \frac{N-1}{N}\sum_{k=1}^N(\xi_i^k - \bar{\xi_i})(\xi_j^k - \bar{\xi_j}).
\end{equation}    
In calculating this error estimate, we ignore any jackknife regions from the analysis with less than 30\% of the median number of groups per jackknife region. We found that the precise criterion adopted
here does not impact the final error estimate. We also tried using sub-regions of size $\sim 220$ sq. degrees and found no noticeable change in the covariance estimates. 

\section{The Measurement and its Interpretation}
\label{sec:measure}

\subsection{Halo Model of the Halo-SZ Cross Correlation}

In order to help interpret the measurements, we use the halo model \citep{Cooray:2002dia} framework to construct theoretical models of the signal. This framework has been applied extensively in the past to study the SZ signal (e.g., \citealt{Komatsu:2002wc,Ostriker:2005ff,Li:2010nk,Fang:2011zk}). Note that the latter two studies explicitly considered the
stacked Compton-$y$ profiles around halos of various masses, as in our current measurements. Here we
briefly describe this model.

The key quantity of interest is 
the average Compton-\y parameter at a projected co-moving distance $r$ from a halo of mass $M$. More
precisely, we are considering the {\em excess} Compton-\y parameter -- above the global-average $y$ -- around
halos of mass $M$.
This can be written as an integral over the halo-pressure cross-correlation function as (e.g. \citealt{Li:2010nk}):
\beqa
\xi_{y,g}(r|M) = \frac{\sigma_T}{m_e c^2} \int_{-\infty}^{\infty} \frac{d\chi}{1+z} \xi_{h, P}(\sqrt{\chi^2 + r^2}|M). \nonumber \\
\label{eq:yprof}
\eeqa
Here $\sigma_T$ is the Thomson scattering cross section, $m_e c^2$ is the rest mass energy of an
electron, $z$ is the redshift of the cluster or group, $\xi_{h,P}$ is the halo-pressure correlation
function, and the integral extends over co-moving coordinates along the line-of-sight (labeled by $\chi$) to the
halo. To simplify notation, we generally suppress redshift labels, although the above quantity clearly depends
on the redshift of the group/halo we are stacking around.

In order to compute the halo-pressure correlation function, we must first determine the gas/electron thermal pressure
as a function of
distance from the halo center (the ``pressure profile'') for halos of various mass. Here we adopt
the fitting formula for the pressure profile from \citet{Battaglia:2011cq}, which is calibrated
from hydrodynamic simulations that include radiative cooling, sub-grid prescriptions for star formation, black
hole accretion, supernova and AGN feedback, and cosmic-ray pressure. In addition to incorporating the impact
of feedback processes, the fitting formula accounts for departures from hydrostatic equilibrium and
for kinetic pressure support. 

\citet{Battaglia:2011cq} use the ``generalized NFW profile'' form to fit the pressure profiles in their simulations, expressing the results in terms of
$x=r/r_{200}$, with
$r_{200}$ denoting the radius at which the average matter density within the halo reaches 200 times the critical density. In order to calculate $r_{200}$ and $M_{200}$ (the
mass enclosed within $r_{200}$) we assume that the overall mass traces a Navarro, Frenk, and White (NFW) profile \citep{Navarro:1996gj}, with the concentration
parameter fit from \citet{Duffy:2008pz}. We assume that halos virialize when the average enclosed density is $\Delta_v(z)$ times the critical density, where
$\Delta_v(z)$ is calculated from the spherical collapse model in LCDM using the fitting formula from \citet{Bryan:1997dn}. The \citet{Battaglia:2011cq} fit is:
\beqa
P_{\rm fit}(x) = P_{200} P_0 (x/x_c)^\gamma \left[1 + (x/x_c)^\alpha\right]^{-\beta}.
\label{eq:pfit}
\eeqa
Here $P_0$, $x_c$, $\gamma$, $\alpha$ and $\beta$ are fitting parameters and $P_{200}$ is the thermal pressure assuming self-similarity:
\beqa
P_{200} = 200 \rho_{\rm cr}(z) \frac{\Omega_b}{\Omega_m} \frac{G M_{200}}{R_{200}}.
\label{eq:pself_sim}
\eeqa
The generalized NFW form is fit to simulated clusters and groups with $M _{200} \geq 5 \times 10^{13} M_\odot$ out to $z=1.5$ (their Figs. 1 \& 2). The
authors fix $\alpha=1.0$ and $\gamma=-0.3$ but tabulate the other fitting parameters (after allowing some redshift dependence for each parameter) in their Table 1.
In the present work, we assume that the fitting formula applies to smaller mass halos, but our results turn out to be insensitive to this assumption.
The electron thermal pressure is related to the total thermal pressure by $P_e(r|M) = \left[(4-2Y)/(8-5Y)\right] P(r|M)$ where $Y$ is the primordial helium mass fraction and
we have assumed that the hydrogen in each halo is highly ionized and the helium doubly-ionized, giving $P_e(r|M) = 0.518 P(r|M)$ for $Y=0.24$. 

We can then proceed to calculate halo-pressure correlation functions and, from that, the Compton-\y profile
around halos of various mass, according to Eq.~\ref{eq:yprof}. The halo-pressure correlation function describes
the average excess pressure around halos -- above that around random points in the universe -- as a function of
the distance from the halo center. This has both a one-halo and a two-halo contribution. The one-halo term describes
the pressure from the hot gas in the halo itself, while the two-halo term corresponds to the contribution from 
correlated neighboring halos. The one-halo term is hence precisely the pressure profile discussed above, with $r$ here denoting the co-moving coordinate separation from the halo center:
\beqa
\xi^{\rm one-halo}_{h,p}(r|M) = P_e(r|M)
\label{eq:xi_oneh}
\eeqa

In order to calculate the two-halo term, it is most convenient to first calculate the two-halo contribution to the
halo-pressure power spectrum and then Fourier transform to find the correlation function. 
Assuming linear-biasing, the two-halo term
is:
\beqa
P_{h,p}(k) = b(M) P_{\rm lin}(k) \int_0^{\infty} dM' \frac{dn}{dM'} b(M') u_P(k|M'). \nonumber \\
\label{eq:two-halo_power}
\eeqa
Here $M$ is the mass of the halo around which we are measuring the Compton-\y parameter, while
$M'$ denotes the mass of a neighbor halo.  The quantity $u_P(k|M')$ is the Fourier transform of the pressure profile around a neighboring halo. Assuming a spherically symmetric pressure profile,
\beqa
u_P(k|M') = \int_0^{\infty} dr 4 \pi r^2 \frac{{\rm sin}(kr)}{kr} P_e(r|M').
\label{eq:press_fourier}
\eeqa
Further, $P_{\rm lin}(k)$ is the linear theory density power spectrum, 
$\frac{dn}{dM'}$ denotes the mass function of the neighboring halos,
while $b(M)$ and $b(M')$ are the linear bias factors of the halos of mass $M$ and $M'$.
We compute the mass function and bias factors using the formulae of \citet{Sheth:2001dp} and \citet{Sheth:1999su}, respectively. Note that the two-halo term includes contributions from neighbors of {\em all masses}, although the
integral is weighted towards fairly massive systems at the redshifts of interest, peaking around $M \sim 5 \times 10^{14} M_\odot$ (see also Fig.~\ref{fig:bp_mass_cont}).

The two-halo contribution to the halo-pressure correlation function follows from Fourier-transforming Equation
\ref{eq:two-halo_power}:
\beqa
\xi^{\rm two-halo}_{h,p}(r|M) = \int_0^{\infty} \frac{dk}{2\pi^2} k^2 \frac{{\rm sin}(kr)}{kr} P_{h,p}(k).
\label{eq:pk_twoh}
\eeqa
The total correlation function is then $\xi_{h,p}(r|M) = \xi^{\rm two-halo}_{h,p}(r|M) + \xi^{\rm one-halo}_{h,p}(r|M)$. Finally, the
average Compton-\y parameter around a halo of mass $M$ comes from integrating the correlation function along the
line of sight, as described by Eqn. \ref{eq:yprof}.

In practice, however, we need to consider the Compton-\y map {\em smoothed} with the Planck beam, and
to weight over the mass distribution and redshift distribution of the groups in each mass bin (\S \ref{sec:scatt_mass}).
We denote the smoothed Compton-\y profile by $\xi^s_{y,g}(\theta|M)$ where $\theta$ is the angle
spanned by the transverse length scale $r$ at redshift $z$, $\theta = r/D_{\rm A, co}(z)$ with $D_{\rm A, co}(z)$ 
denoting the co-moving angular diameter distance. We
find the smoothed profile by calculating the angular cross power spectrum, $C_{x, \sc l}$, (in the flat sky approximation), as
\beqa
C_{x, \sc l} = \int d\theta 2 \pi \theta J_0({\sc l} \theta) \xi_{y,g}(\theta|M), 
\label{eq:cl_cross}
\eeqa
where $J_0({\sc l} \theta)$ is a zeroth-order Bessel function. We then
multiply by the Planck beam, and Fourier-transform again, to find the smoothed correlation function:
\beqa
\xi^s_{y,g}(\theta|M) = \int \frac{d{\sc l} {\sc l}}{2 \pi} J_0({\sc l} \theta) C_{x, \sc l} B_{\sc l}.
\label{eq:yprof_smooth}
\eeqa
$B_{\sc l}$ gives the Planck beam profile in Fourier space, 
$B_{\sc l} = {\rm exp}[-{\sc l}({\sc l}+1) \sigma^2/2]$ and $\sigma = \theta_{\rm FWHM}/\sqrt{8 {\rm ln(2)}}$. Here $\theta_{\rm FWHM}$ is the full-width
at half-maximum of the instrumental beam, which we set to $10$ arcmin since the Planck \y maps are smoothed at this resolution (see \ref{sec:ymap}).

It is also interesting to consider the large-scale limit, $r >> r_{\rm vir}$, in which case we can
take the $k \rightarrow 0$ limit of $u_P(k|M)$, where $u_p \rightarrow {\rm const}$.  
In this limit, we get
\beqa
\xi^s_{y,g}(r >> r_{\rm vir}|M) \approx \frac{\sigma_T}{m_e c^2} \frac{w^S_{\rm lin}(r)}{1+z} b(M) \avg{b P_e}. \nonumber \\
\label{eq:yprof_large_scale}
\eeqa
Here we have assumed linear biasing, and $w^S_{\rm lin}(r)$ is the projected linear matter correlation function
(with units of co-moving length) smoothed by the Planck beam, and
 $\avg{b P_e}$ denotes the average bias-weighted thermal energy (per unit proper volume) -- i.e.,
the average bias-weighted electron pressure -- of the universe:
\beqa
\avg{b P_e}(z) = \int dM' \frac{dn}{dM'} (1+z)^3 E_T(M') b(M'),
\label{eq:bias_pressure}
\eeqa
with $E_T(M')$ denoting the thermal energy of the electrons in 
a halo of mass $M'$, which is the integral of the electron pressure
profile over proper volume. 
Eq.~\ref{eq:yprof_large_scale} implies that we can combine the large scale
stacked Compton-\y parameter measurements in our different group mass bins to best estimate $\avg{b P_e}$, although because of their differing redshift distributions each group mass bin probes
a slightly different redshift-weighted average. We describe this in more
detail below in Eq.~\ref{eq:bp_estimate} below.

\subsection{Scatter in the Relationship Between Group Mass and Halo Mass}
\label{sec:scatt_mass}

So far we have discussed modeling the SZ-{\em halo} mass correlation function. In order to compare with our measurements, we need to further account for:
scatter in the relationship between \citet{yang07}'s halo mass estimate and the true halo mass; for the mass extent of our different group mass bins; and for
the redshift distribution of the groups within each mass bin. We refer to \citet{yang07}'s halo mass estimate as the ``group mass''.

In order to account for scatter in the halo mass estimate, we take a forward modeling approach (along the lines of \citealt{Lima:2005tt}). We start by assuming that the
\citet{yang07} group mass, $M_{\rm obs}$, is correct on average but that there is some scatter in this estimate around the true halo mass, $M_{\rm true}$. According
to \citet{yang07}, the scatter is well described by a lognormal distribution with a dispersion of around $0.2-0.35$ dex, with the precise value depending on the group
mass bin. In this case,
\beqa
\frac{dP(M_{\rm obs}|M_{\rm true})}{d\rm{ln}M_{\rm obs}} = \frac{1}{\sqrt{2 \pi \sigma^2_{\rm ln M}}} \rm{exp} \left[-x^2\right],
\label{eq:logn}
\eeqa
with
\beqa
x = \frac{\rm{ln}(M_{\rm obs}) - \rm{ln}(M_{\rm true})}{\sqrt{2 \sigma^2_{\rm ln M}}},
\label{eq:zfac}
\eeqa
i.e., ${dP(M_{\rm obs}|M_{\rm true})}{d\rm{ln}M_{\rm obs}}$ describes the differential probability distribution that the group mass is estimated to be $M_{\rm obs}$ when the true halo mass is actually
$M_{\rm true}$. We can then consider the average abundance of groups in a bin with mass between $M_{\rm obs} = M_L$ and $M_{\rm obs} = M_H$ at redshift $z$ . We will later account for the redshift distribution of groups
in each mass bin. The abundance is:
\beqa
n_{\rm obs} =  \int_{M_L}^{M_H} \frac{dM_{\rm obs}}{M_{\rm obs}} \int_0^\infty \frac{dM_{\rm true}}{M_{\rm true}} \frac{dn}{d\rm{ln} M_{\rm true}}  \frac{dP(M_{\rm obs}|M_{\rm true})}{d\rm{ln}M_{\rm obs}}. \nonumber \\
\label{eq:nobs}
\eeqa
Here $d\rm{ln}/dM_{\rm true}$ is the underlying halo mass function.
It is useful to interchange the order of integration and carry out the integral over $M_{\rm obs}$ analytically, as in \citet{Lima:2005tt}. Then we only need to evaluate the integral over $M_{\rm true}$:
\beqa
n_{\rm obs} =  \int_0^\infty \frac{dM_{\rm true}}{M_{\rm true}} \frac{dn}{d\rm{ln} M_{\rm true}}  
\frac{1}{2} \left[{\rm erfc}(x_L) - {\rm erfc}(x_H)\right], \nonumber \\
\label{eq:nobs_single}
\eeqa
Here $x_L$ is the argument of the lognormal above (Eq.~\ref{eq:zfac}) evaluated at the lower end of the mass bin, $M_{\rm obs} = M_L$, and $x_H$ is the same evaluated at the upper end of the mass bin, $M_{\rm obs} = M_H$.
We can then evaluate Eq.~\ref{eq:nobs_single} using the \citet{Sheth:2001dp} model for the halo mass function $dn/d\rm{ln} M_{\rm true}$. 
Similarly, the smoothed \y-group correlation function in the mass-bin of interest can be expressed as a weighted-average over the true halo mass function as:
\begin{align}
\xi^s_{y,g}(r,z) =& \frac{1}{n_{\rm obs}(z)} \int_0^{\infty} \frac{dM_{\rm true}}{M_{\rm true}} \frac{dn(z)}{d\rm{ln} M_{\rm true}} \xi^s_{y,g}(r|M_{\rm true};z) \nonumber \\
& \times \frac{1}{2} \left[{\rm erfc}(x_L) - {\rm erfc}(x_H)\right].
\label{eq:master}
\end{align}
The above equation hence gives the (smoothed) excess \y at a co-moving separation $r$ from groups in a mass bin between $M_L$ and $M_H$ at redshift $z$. In this equation we have, for clarity, restored the redshift labels $z$ that we generally suppress. Finally, recall that in each mass bin we combine measurements in narrow redshift slices (Eq.~\ref{eq:yest}) to reduce the noise in our estimates.
The final step in our modeling is therefore to weight by the probability, $dP/dz$, that a group in the mass bin lies within a given redshift range (Fig.~\ref{fig:zdist-group}):
\beqa
\xi^s_{y,g}(r) = \int_0^{\infty} dz \frac{dP}{dz} \xi^s_{y,g}(r,z).
\label{eq:zweight}
\eeqa

Along the lines of Eq.~\ref{eq:yprof_large_scale}, we can also take the large-scale limit of Eq.~\ref{eq:zweight} and combine redshift slices to estimate
the bias-weighted pressure at the redshifts probed in each group mass bin:
\beqa
\avg{b P_e} = \frac{m_e c^2}{\sigma_T} \int_0^{\infty} dz \frac{dP}{dz}\frac{1+z}{w^s_{\rm lin}(r,z)} \frac{\xi^s_{y,g}(r >> r_{\rm vir},z)}{\avg{b_{\rm group}(z)}}, \nonumber \\
\label{eq:bp_estimate}
\eeqa
where $\avg{b_{\rm group}(z)}$ is the average bias of the groups in the mass bin of interest for the narrow redshift slice centered on $z$.  In principle, we could measure $\avg{b_{\rm group}}(z)$ from
the auto-correlation function of the groups in each mass bin and redshift slice. Here we instead model this, using Eq.~\ref{eq:master}, except with $\xi^s_{y,g}(r|M_{\rm true};z)$ replaced
by $b(M_{\rm true};z)$.  Since each group mass bin has a different redshift distribution (Fig.~\ref{fig:zdist-group}), we will get slightly different results for $\avg{b P_e}$; each mass bin is measuring
the average bias-weighted pressure at slightly different redshifts. Nevertheless, Eq.~\ref{eq:bp_estimate} provides a useful and compact description of the two-halo term and it's information content.
With these equations in hand, we predict $\xi^s_{y,g}(r)$ and $\avg{b P_e}$ for each mass-bin in our sample.

\subsection{The Measured SZ-Group Cross Correlation}
\label{sec:twop_results}

\begin{figure*}
\centering
\includegraphics[scale=0.7]{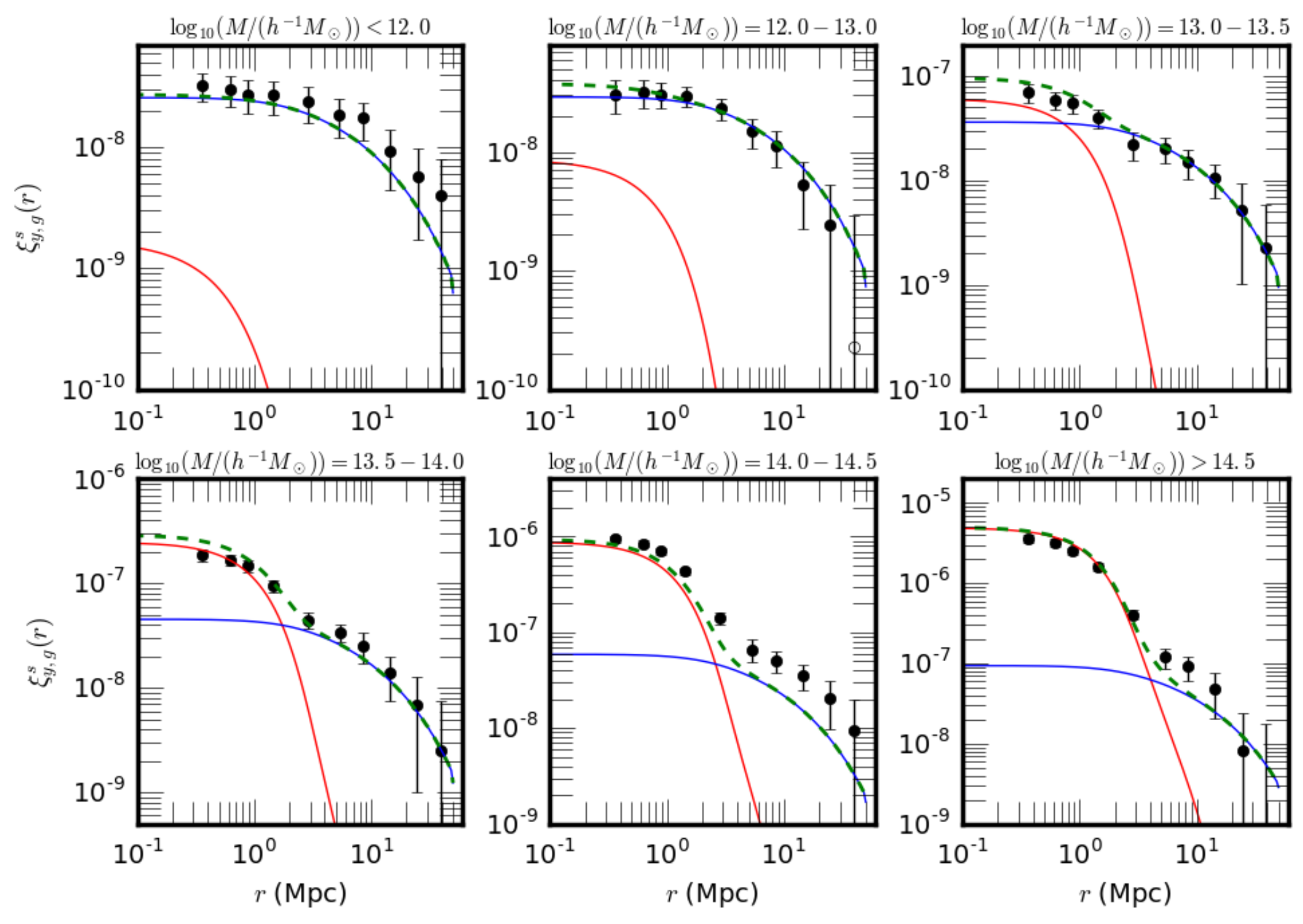}
\caption{ The SZ-group correlation function, as a function of co-moving separation $r$, for each of six bins in group mass. 
The black solid circles show the measurement, made according to Eq. \ref{eq:yest}. 
The error bars are from jackknife estimates of the diagonal part of the covariance matrix, but note that the radial and mass bins are correlated here (See Fig. \ref{fig:pm-xi-cov}).
In each panel, the green dashed line shows the fiducial halo model (after smoothing by
the Planck beam), assuming the \citet{Battaglia:2011cq} pressure profile. The model curves adopt our fiducial assumptions for mis-centering, and scatter and bias in the group-mass halo-mass relation (see text). The red and blue curves in each panel show, respectively, the one and two-halo contributions to
the two-point function. 
}
\label{fig:pm-xi}
\end{figure*}

Now that we have described our approach for estimating the two-point function and the halo model used to interpret the measurements, we can start to examine the results.
The main result of this paper is Fig. \ref{fig:pm-xi} which shows the full SZ-group two point correlation function as a function of co-moving radius, in each of six bins
in group mass. The results here use the Planck MILCA \y-map.
The black points show the measurement of the two-point correlation function according to Eq. \ref{eq:yest}.

The results shown in the figure suggest that we have statistically significant detections in each of the separate group mass bins, as we quantify further below. The measurements are compared to
halo model predictions, as described in the previous section, based on the \citet{Battaglia:2011cq} simulated pressure profiles. As we will discuss, we find that it is important to account
for small systematic errors in the SDSS group catalog. Specifically, we need to allow mis-centering errors (i.e., each SDSS group does not in fact lie exactly at the center of its host dark matter
halo). In addition, we find that allowing slightly more scatter in the group mass-halo mass relation than suggested in \citet{yang07} and a small average bias in this relation generally improves
agreement between the models and data. Although the group-mass halo-mass relation is calibrated
in the analysis of \citet{yang07}, this relies on abundance-matching and mock catalogs which may be imperfect. Indeed, even small errors in this relation may potentially be important here.

Quantitatively, \citet{yang07} find that the group-mass halo-mass relation has a scatter of 0.35 dex between $10^{13} h^{-1} M_\odot < M < 10^{14.5} h^{-1} M_\odot$ and a scatter of 0.2 dex at lower and higher masses. In our baseline model -- which we refer to as our ``fiducial model'' in what follows -- we assume slightly larger values of the scatter, adopting 0.25 dex
in our two lowest mass bins, 0.40 dex in the next three bins, and 0.30 dex in the highest mass bin. Therefore, our fiducial model assumes a higher scatter than in \citet{yang07} by 0.05 dex, except
in our highest mass bin where we adopt a still larger value. The lower end of this bin ($10^{14.5} h^{-1} M_\odot$) coincides with the end of the mass range where these authors find a higher scatter and so the appropriate scatter is a bit ambiguous for this bin. Our fiducial model additionally incorporates a $10\%$ bias, assuming that the halo-mass estimate is a slight overestimate on average, by setting 
${\rm ln}(M_{\rm obs}) \rightarrow {\rm ln}(M_{\rm obs}) + {\rm ln}(0.9)$ in Eq.~\ref{eq:zfac}. 

Unfortunately, it is unclear exactly how large the mis-centering effect may be for the \citet{yang07} SDSS groups. One possible empirical guide comes from the analysis
of \citet{Johnston07}, who used galaxy-galaxy lensing measurements to constrain the offset of brightest cluster galaxies (BCGs) in the maxBCG catalog from their halo centers.
They found that a fraction $p_c$ of these galaxies reside in their halo centers, while $1-p_c$ are offset and that the offset is well described by a 2D Gaussian distribution. The distribution of spatial offsets is centered around zero, but with a significant rms dispersion of $0.42 h^{-1}$ proper Mpc. It is not at all clear that a similar offset distribution should apply to the \citet{yang07} catalog. For one,
the \citet{yang07} catalog centroids based on the entire set of member galaxies, rather than on the brightest central galaxy. There are numerous other differences in the mass range spanned and algorithms employed in the maxBCG and \citet{yang07} catalogs. Our fiducial model nevertheless adopts the mis-centering distribution found by \citet{Johnston07}, assuming that it applies to the SDSS group catalog
here. We adopt the centering fractions from their analysis, based on the average mass in each of our bins, which yields $p_c = 0.53, 0.54, 0.58, 0.63, 0.72,$ and $0.83$, in order of increasing mass.
Note that the rms dispersion is large relative to the virial radius in the low mass bins, but the
mis-centering effect is important mostly in the higher mass bins (where it is a small fraction of the virial radius), as we will see. For example, it is larger than the virial radius in the lowest mass bin, and 0.69 times $R_{200}$ for our second bin, while it is $0.15$ times $R_{200}$ in the highest mass bin.
The mis-centering is incorporated by (further) smoothing the model two-point
function according to the distribution above for a fraction $1-p_c$ of the groups.
In this work, we fix the mis-centering, scatter, and bias parameters
to plausible values, rather than performing a full marginalization. Although our fiducial choices here are somewhat arbitrary, we consider variations around them in the next sub-section.

In the three most massive bins with $M \geq 10^{13.5} h^{-1} M_\odot$, we see clear evidence for both a one-halo term (specified by the solid red line in the models) and a two-halo contribution (solid blue lines). On small scales, our measurements in these bins  probe the hot gas distribution inside the groups, while at large separations we pick up the cumulative contribution from hot gas in correlated neighboring halos. 
In the three smallest group mass bins, we expect the two-halo contribution to be more important at small separations. In fact, in the two smallest mass bins, the two-halo term should completely dominate
over the one-halo contribution on all measurable scales. Indeed, in the three lowest mass bins the measurements appear almost entirely consistent with the two-halo term alone (the only exceptions are the inner few radial points in the $M=10^{13}-10^{13.5} h^{-1} M_\odot$ bin). 
This implies that these measurements {\em are not probing the distribution of hot gas around the low mass systems, but instead reflect the contribution from more massive correlated neighboring systems. }  
To detect a possible one-halo contribution from these smaller mass groups, we will need data with better resolution than  Planck (see \S \ref{sec:future}). 
The dominant
contribution to the two halo term comes from halos near $M \sim 10^{14.5} M_\odot$ (given the redshifts of our current group sample). The lower mass systems then effectively just provide
additional ``stacking centers'' that we can use to boost our sensitivity to the two-halo term.

\begin{figure*}
\centering
\includegraphics[scale=0.6]{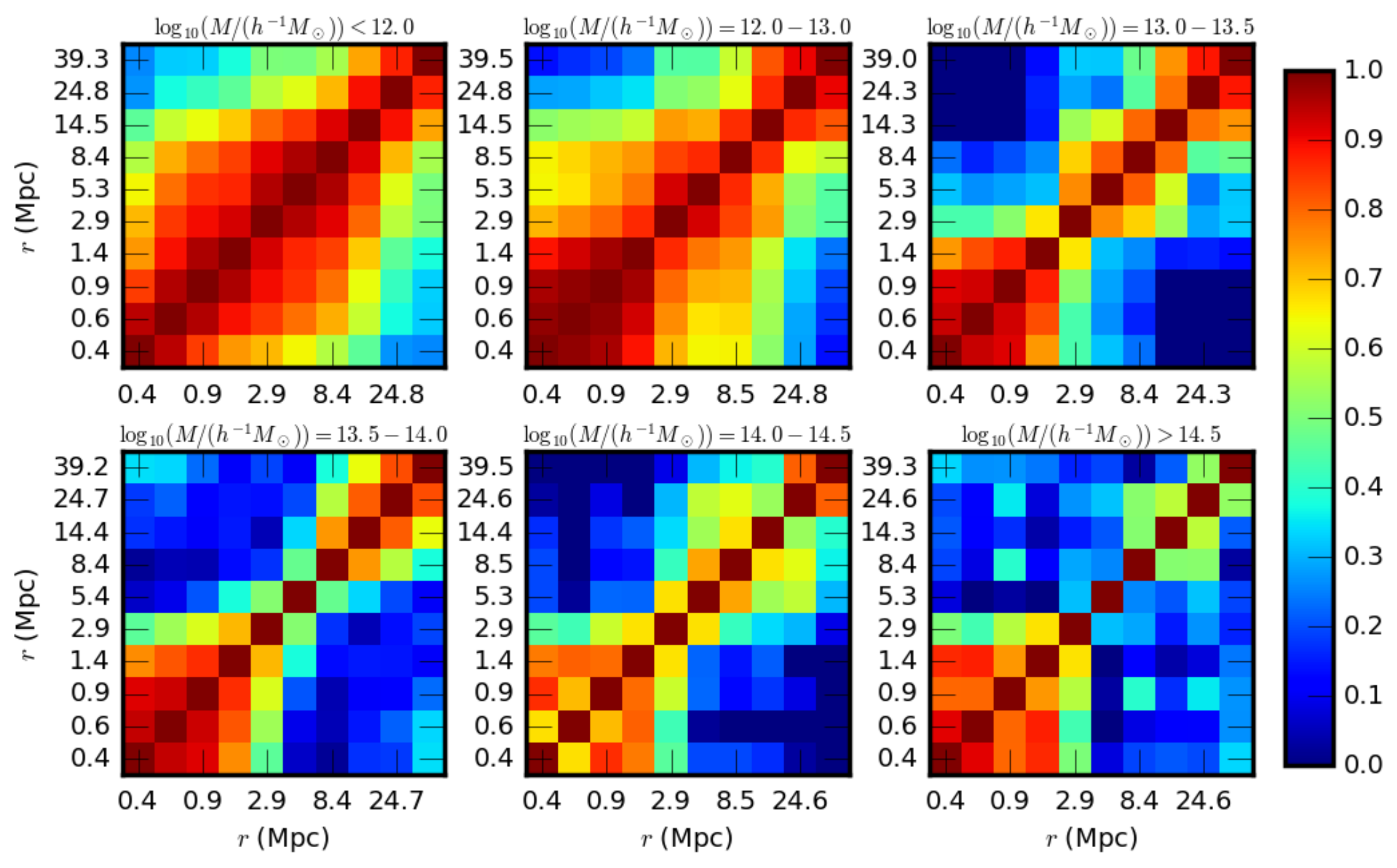}
\caption{Covariance matrices of $\xi^s_{y,g}(r)$ for each of the six group mass bins. The color bar gives the cross-correlation coefficient between estimates of the two-point function $\xi^s_{y,g}$ in different radial bins,
$r_i$ and $r_j$. Here $r_i$ and $r_j$ vary along the x and y-axes of each panel and the cross-correlation coefficient is shown for each pair of radial bins. 
} 
\label{fig:pm-xi-cov}
\end{figure*}

In order to make more quantitative statements, we must invert the full covariance matrix and calculate the chi-square goodness-of-fit between the model and the data.
Our jackknife estimate of
the covariance matrix of $\xi^s_{y,g}(r)$ between different radial bins is shown (for each halo mass bin) in Fig. \ref{fig:pm-xi-cov}. In the two lowest mass bins, there are strong correlations between
neighboring radial bins while the more massive bins are closer to diagonal. In the case of the low mass systems, the clustering of the host halos is important and hence measurements in different
radial bins often involve stacking nearly the same patch of \y map, leading to these off-diagonal correlations. In the higher mass systems, however, Poisson fluctuations dominate over the clustering term, and this leads to smaller off-diagonal correlations. Although the overall form of the covariance matrices looks sensible, our jackknife estimates are somewhat noisy and this makes
it difficult to robustly estimate chi-square values. In order to handle this, we regularize the covariance matrix by removing noisy eigenvectors with small eigenvalues and inverting the matrix with these
eigenvectors truncated. Specifically, we find that our results are stable after truncating eigenvalues that are a factor of 0.01 smaller than the sum of the matrix's eigenvalues.

We first assess the overall significance of the SZ-group correlation functions by comparing our measurements to a null model. The results of these calculations are given in the first column of Table \ref{tab:significance}. In each mass bin, the SZ-group correlation is measured at high significance. The weakest measurement is in the lowest mass bin (with $M \leq 10^{12} h^{-1} M_\odot$), but even this is a strong $5.7-\sigma$ detection.  Our most significant detection is a $15-\sigma$ measurement in the $10^{14} -10^{14.5} h^{-1} M_\odot$ bin. 
(The second column of Table \ref{tab:significance} quantifies the results of a null test -- this is discussed later in \S \ref{sec:systematics}.)
Next, we compute the
goodness-of-fit for the baseline models of Fig.~\ref{fig:pm-xi}. The first column of Table \ref{tab:groups-chi2nu} gives the value of chi-square per degree of freedom, $\chi^2_\nu$, in each mass bin.
In the first three mass bins, the models are acceptable fits with $\chi^2_\nu \sim 0.9-1.3$, while the three higher mass bins are poorer fits. In the four highest mass bins -- with the exception of
the $10^{14}-10^{14.5} h^{-1} M_\odot$ bin -- the model curves are larger than the data in the one-halo dominated regime. On the other hand, the $10^{14}-10^{14.5} h^{-1} M_\odot$ model
is smaller than the data in both the one and two-halo dominated regimes. In all of the other mass bins, the model two-halo term provides a fairly good fit to the measurements in the regime
where it dominates. 

\begin{table}
\begin{center}
 \begin{tabular}{ccc}
$\log_{10}(M/(h^{-1} M_\odot))$ & S/N & $\chi^2_{\nu,null}$ \\
\hline
$<$ 12.0 & 5.7 & 0.4 \\
12.0-13.0 & 7.2 & 1.1  \\
13.0-13.5 & 6.7 & 0.8\\
13.5-14.0 & 10.9 & 0.9 \\
14.0-14.5 & 15.0 & 0.4 \\
$>$ 14.5 & 13.3 & 1.0 \\
\end{tabular}
\end{center}

\caption{Significance of SZ-group detection and null test. The first column shows the overall detection significance (i.e., the $\sigma$-level) of the SZ-group correlation signal in each mass bin.
The second column gives $\chi^2$ per degree of freedom for our null test (see text, \S \ref{sec:systematics}).  
\label{tab:significance}}
\end{table}

\begin{table*}
\begin{center}
 \begin{tabular}{cccccc}
$\log_{10}(M/(h^{-1} M_\odot))$ & Fiducial & Simple & Mis-centered & Large scatter & Mass bias \\
\hline
$<$ 12.0 & 1.3 & 1.2 & 1.3 & 1.2 & 1.3 \\
12.0-13.0 & 1.1 & 1.5 & 0.9 & 1.4 & 1.2 \\
13.0-13.5 & 0.9 & 3.5 & 0.7 & 1.6 & 1.2 \\
13.5-14.0 & 1.9 & 13.0 & 1.9 & 3.2 & 3.3 \\
14.0-14.5 & 2.4 & 6.7 & 2.0 & 4.6 & 4.2 \\
$>$ 14.5 & 3.7 & 23.7 & 16.8 & 3.3 & 6.9 \\
\end{tabular}
\end{center}
\caption{Goodness-of-fit for different models. Each entry shows the value of $\chi^2_\nu$ (chi-square per degree of freedom) for one of our models and mass bins, with the different models making
varying assumptions about systematics in the SDSS group catalog. The first column gives $\chi^2_\nu$ for
our fiducial mis-centering, and mass scatter/bias parameters.  The other columns show $\chi^2_\nu$ for  the ``Simple model'',  ``Mis-centered model'', ``Large scatter model'' and ``Mass bias model'', respectively. 
}
\label{tab:groups-chi2nu}
\end{table*}

\subsection{Group Catalog Systematics}
\label{sec:group_systematics}

It is possible that the discrepancies observed between the models and measurements reflect interesting differences between the hot gas distribution in the observed groups/clusters and the
simulated \citet{2012ApJ...758...74B} pressure profile. Alternatively, they might indicate a bias in our measurement or reflect systematics in the MILCA \y map, as we will further investigate
subsequently (\S \ref{sec:systematics}). However, another possibility
is that the differences instead (mostly) reflect remaining systematic errors in the group catalog. Although we defer a more complete treatment to future work, here we consider the impact
of variations around our fiducial mis-centering errors, and mass scatter/bias parameters. Note that each of these effects tends to lower the model one-halo term, without significantly impacting
the two-halo term and so these effects may be responsible for differences between the models and measurements in inner radial bins, but can not account for discrepancies at large radius, such
as those seen most prominently in the $M=10^{14}-10^{14.5} h^{-1} M_\odot$ mass bin (Fig.~\ref{fig:pm-xi}).

\begin{figure*}
\begin{center}
\includegraphics[scale=0.7]{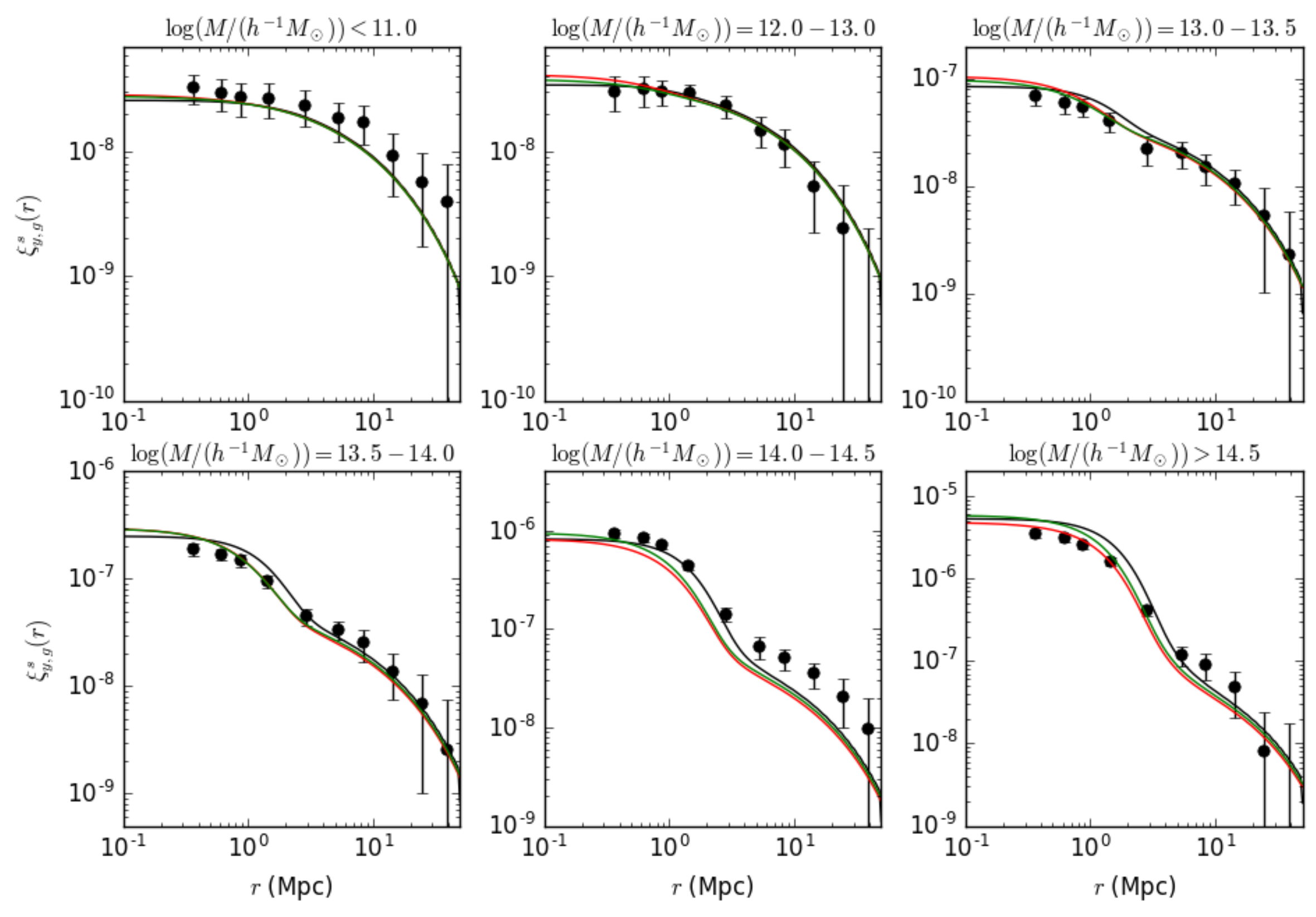}
\caption{Impact of mis-centering, bias, and increased scatter in the group-mass halo-mass relation. Here the measured group-SZ correlation function is compared to models which incorporate these
effects (only the sum of the one and two halo terms is shown for each model). The yellow, green, and blue lines show, respectively, the impact of mis-centering, increased mass scatter, and mass bias. 
Each of these effects lowers the model one-halo term somewhat, and improves the agreement with the data. For comparison, the black and red lines show the fiducial and simple models, respectively.
}
\label{fig:group-vars}
\end{center}
\end{figure*}

Fig.~\ref{fig:group-vars} and Table \ref{tab:groups-chi2nu} show the result of varying our group catalog systematics parameters around their fiducial values. The red line in Fig.~\ref{fig:group-vars}, which
we denote as our ``simple model'', includes scatter at the level of \citet{yang07} -- i.e., 0.2 dex in the lowest two and highest group mass bins, and 0.35 dex in the other three bins -- with no mis-centering
and no mass bias. This model is a significantly poorer fit than our fiducial case, except in the lowest mass bin where the group systematics parameters have only a small effect. The simple model overpredicts the measurements in the one-halo regime for almost every mass bin. 

In order to isolate the effect of mis-centering, we consider a model with the same rms mis-centering error (from \citealt{Johnston07}) as our fiducial case, but instead assume the extreme scenario of
$p_c=0$ -- i.e., that the mis-centering error applies to all groups in the catalog -- without allowing any mass bias and taking scatter at the level of \citet{yang07}. 
 The effect of this additional smoothing may be discerned most cleanly by comparing the red and yellow solid curves in Fig.~\ref{fig:group-vars}. In this extreme case, mis-centering errors alone would mostly suffice to explain the differences between the model and data in the inner radial bins, although the model in the most massive bin still overproduces the measurement. (See also Table \ref{tab:groups-chi2nu}.) In the more realistic case that $p_c$ is larger, mis-centering can not provide the full explanation for the differences. 

In order to better understand the impact of the group-mass halo-mass relation, we consider first the case of increasing the 
scatter by 0.15 dex in each group mass bin. In this case, we adopt a scatter of 0.35 dex in the
lowest two group mass bins and in the highest mass bin, while we assume a scatter of 0.50 dex in the other three bins. The increased scatter makes the largest difference in the highest mass bin. This is because the mass-function is very steep at the high mass end, and increasing the scatter mostly brings low mass halos into the bin and this results in a lower average Compton-\y signal. In the highest
mass bin, the increased scatter can explain most of the discrepancies between the model and data. 
Finally, we consider the impact of a larger $30\%$ bias -- specifically an overestimate -- in the group-mass estimate by setting ${\rm ln}(M_{\rm obs}) \rightarrow {\rm ln}(M_{\rm obs}) + {\rm ln}(0.7)$ in Eq.~\ref{eq:zfac}. Again, we find that this bias could explain some of the discrepancies in the one-halo regime: for this to be the sole culprit, however, requires an uncomfortably large mass bias. These
differences may be discerned visually in Fig.~\ref{fig:group-vars}, while the quantitative improvements are given in Table \ref{tab:groups-chi2nu}.

Although rather extreme assumptions are required for any one of these effects to explain the differences between the simple model and the measurements alone, they likely act together. Our fiducial model is meant to adopt reasonable values for each of the mis-centering, mass scatter, and bias parameters. 
The examples of this sub-section illustrate the importance of accurately characterizing the group-mass halo-mass relation for this analysis. Future efforts aimed at more precise measurements
of the one-halo term around low mass systems will require improvements here. One promising approach is to calibrate these relations using galaxy-galaxy lensing measurements. 

\subsection{Estimation of $\langle b P_e \rangle$}

Following Eq.~\ref{eq:bp_estimate}, we can combine the measurements from each group mass bin in the two-halo dominated regime to estimate the bias-weighted electron pressure at the redshifts of the groups in our sample. To do this, we use the measurements of the correlation function between radii of $5-25$ co-moving Mpc, where the two-halo term should dominate. We found that our results
are insensitive to the precise choice of inner and outer radius, as long as the inner radius is larger than about $3$ co-moving Mpc. Using the (smoothed) correlation function from linear theory and a \citet{Sheth:1999su}
based model for the bias of each group mass bin (after accounting for scatter in the group-mass halo-mass relation), we invert a subsection of the covariance matrix of interest -- within the radial range specified -- to find
$\langle b P_e \rangle$. Strictly speaking, we expect slightly different results for each mass bin because of the differing redshift distribution of the groups in each bin. However, this
spread is expected to be small compared to the statistical errors on our present measurements. Based on the \citet{Battaglia:2011cq} pressure profile, we expect $\avg{b P_e} = 1.6 \times 10^{-7}$ keV cm$^{-3}$ after averaging over the redshift distribution in the lowest
mass bin, and $\avg{b P_e} = 1.8 \times 10^{-7}$ keV cm$^{-3}$ using the redshift distribution in the highest mass bin. 

\begin{figure}
\begin{center}
\includegraphics[width=\columnwidth]{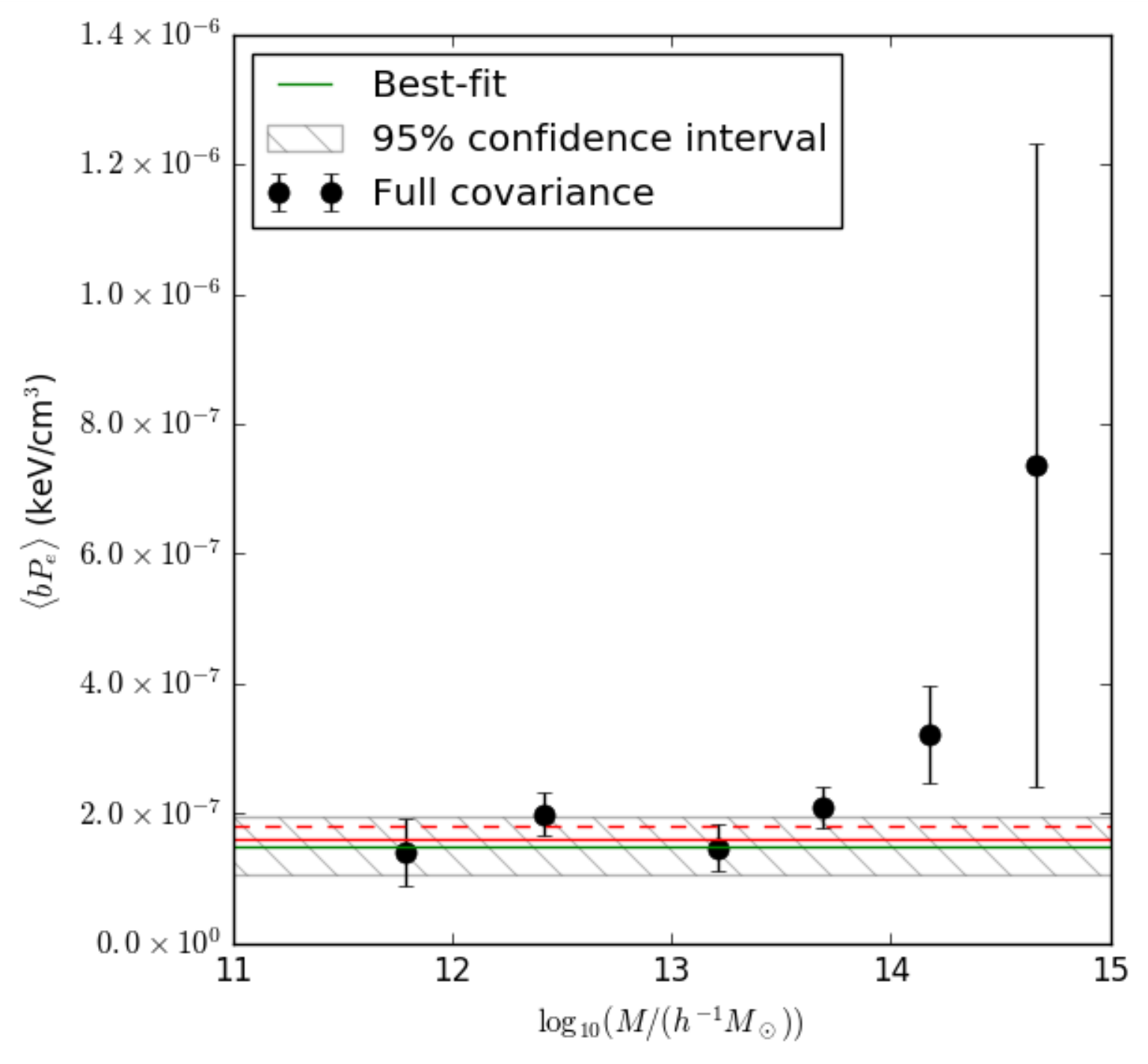}
\caption{Measurements of the bias-weighted electron pressure at the redshifts of the SDSS groups. The points show separate estimates of $\avg{b P_e}$ from each mass bin, while the solid and dashed red lines show the theoretical expectation; the solid (dashed) line shows the model prediction given the redshift distribution of the groups in the lowest mass (highest mass) bin. The redshift distribution of the groups in the lowest (highest) mass bin is the most concentrated of the mass bins towards low (high) redshift, so the model for each mass bin lies within the range specified by these two lines. The best-fit
value after combining all of the mass bins is the green solid line, while the hatched region shows the corresponding 95\% confidence interval.  
}
\label{fig:bias-pressure}
\end{center}
\end{figure}

The results of these measurements are shown in Fig.~\ref{fig:bias-pressure}. As expected from Fig.~\ref{fig:pm-xi}, the measurements mostly agree with the model predictions except in the $10^{14} h^{-1} M_\odot < M
< 10^{14.5} h^{-1} M_\odot$ bin, in which the data points are higher than the model by $2.3-\sigma$. Aside from this single discrepant bin, the measurements of $\avg{b P_e}$ are
consistent across the different mass bins.
Ignoring the small variations from the differing redshift distributions, we can then
combine the measurements from all of the mass bins to best estimate $\avg{b P_e}$ across the entire data sample. In this estimate, it is important to account for correlations between the measurements in the different mass bins. Specifically, we find that the correlation coefficients between our $\avg{b P_e}$ estimates in different mass bins are typically larger than $0.5$. 
Accounting for these correlations, our best estimate of the average bias-weighted pressure across all group mass bins is $\langle b P_e \rangle = 1.50 \times 10^{-7} \pm 2.26 \times 10^{-8}$ keV cm$^{-3}$. 
This differs from the model expectation at only $\sim 1-\sigma$, and so the measurement is consistent with theoretical expectations 
If we further assume the model to divide out the pressure-weighted average bias of $\avg{b} = 2.9$, we can infer the total average electron pressure of the universe near $z \sim 0.1-0.2$ to be $\avg{P_e} = 5.2 \pm 0.79 \times 10^{-8}$ keV cm$^{-3}$. 
This is an interesting result: we have obtained a measurement of the total average thermal energy content of the universe at the redshifts of the SDSS group sample.

\section{Further Tests for Systematic Errors}
\label{sec:systematics}

\begin{figure}
\begin{center}
\includegraphics[width=\columnwidth]{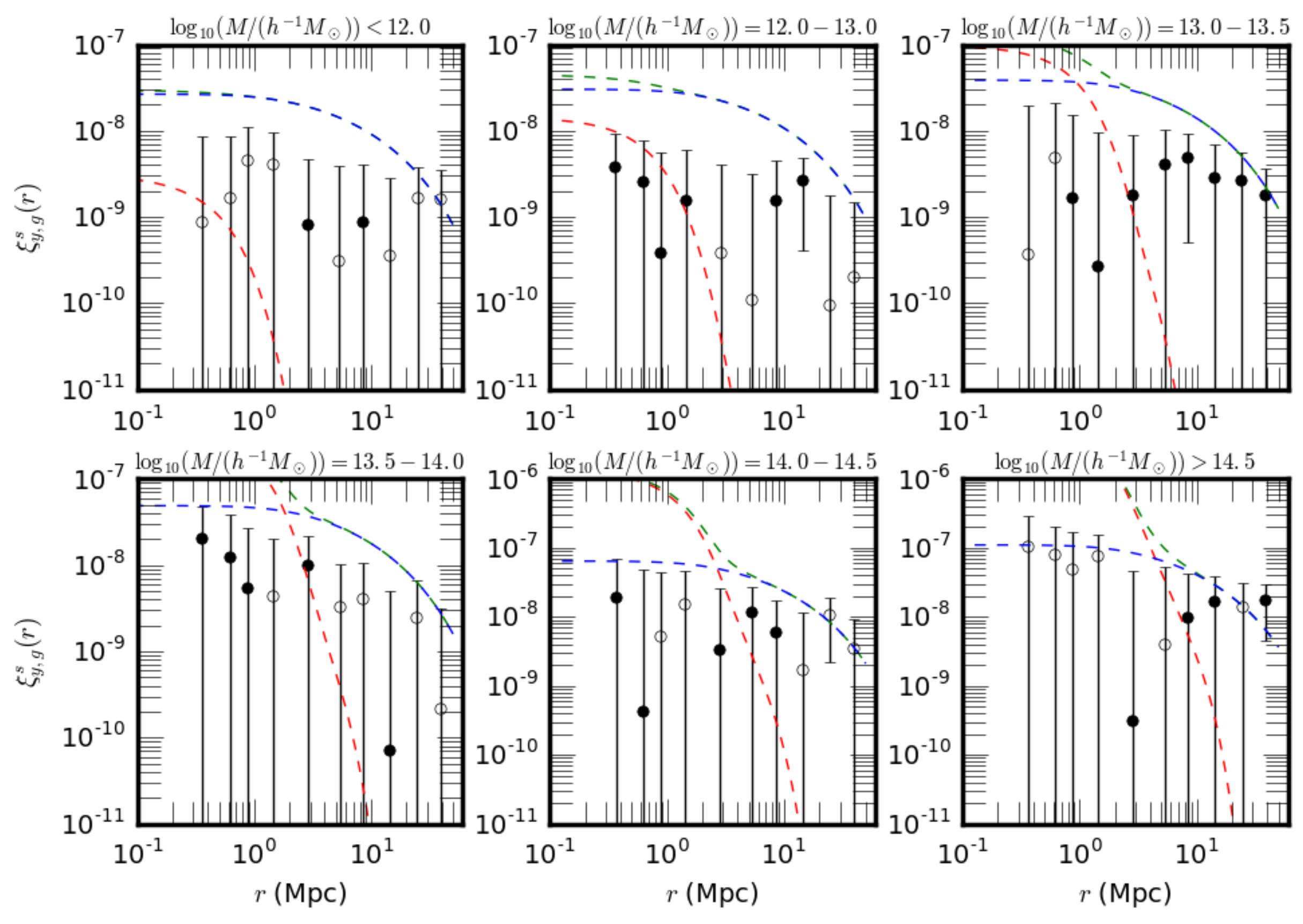}
\caption{Systematics test using the MILCA null map.  Here we stack the MILCA null map (see text) around the SDSS groups in each mass bin. In the absence of systematic errors,
we expect the null map-group correlation function to be consistent with zero to within the statistical error bars of the measurement. We find that the null-group correlation function is consistent
with zero for each mass bin (i.e., $\chi^2$ per degree of freedom is close to one in each case).}
\label{fig:null-milca}
\end{center}
\end{figure}

In order to further assess the reliability of our results, we perform several additional tests for systematic errors. The first test utilizes the so-called ``First Minus Last'' MILCA map, which we refer to as the MILCA null map. The Planck collaboration constructed this map by first dividing the 
data into two separate time chunks, performing the \y reconstruction on each chunk, and then differencing the two estimates. The resulting difference map should entirely remove the Compton-\y signal --
which is present in each of the two time chunks -- while leaving noise that is uncorrelated across the two separate time splits. We can then correlate this difference map with the SDSS groups, using exactly
the same estimator (Eq.~\ref{eq:yest}) and analysis as in our main measurement (Fig.~\ref{fig:pm-xi}). In the absence of systematic errors, this measurement should be consistent with zero -- i.e.,
the SDSS groups should not correlate with noise in the Planck \y maps. The results of this test are shown in Fig.~\ref{fig:null-milca} for the case of the MILCA \y map, and the corresponding 
values of $\chi^2_{\nu, null}$ are given in Table \ref{tab:significance}. Reassuringly, these measurements are consistent with zero: our method is not producing
a spurious correlation in this case (where correlations are unexpected.)  However, some of the values of $\chi^2$ per degree of freedom are a bit small, with $\chi^2_\nu = 0.4$ in the lowest mass bin
and in the $10^{14} h^{-1} M_\odot < M < 10^{14.5} h^{-1} M_\odot$ bin. Since $\nu=10$, the value of $\chi^2$ itself is $\chi^2=4$, while the expected value is $\avg{\chi^2} = 10$ with a standard deviation of $\sqrt{2 \nu} = 4.5$. Hence, $\chi^2_\nu=0.4$ not unreasonably low, but this could be a statistical fluke or it may indicate the error bars are somewhat overestimated.

\begin{figure}
\begin{center}
\includegraphics[width=\columnwidth]{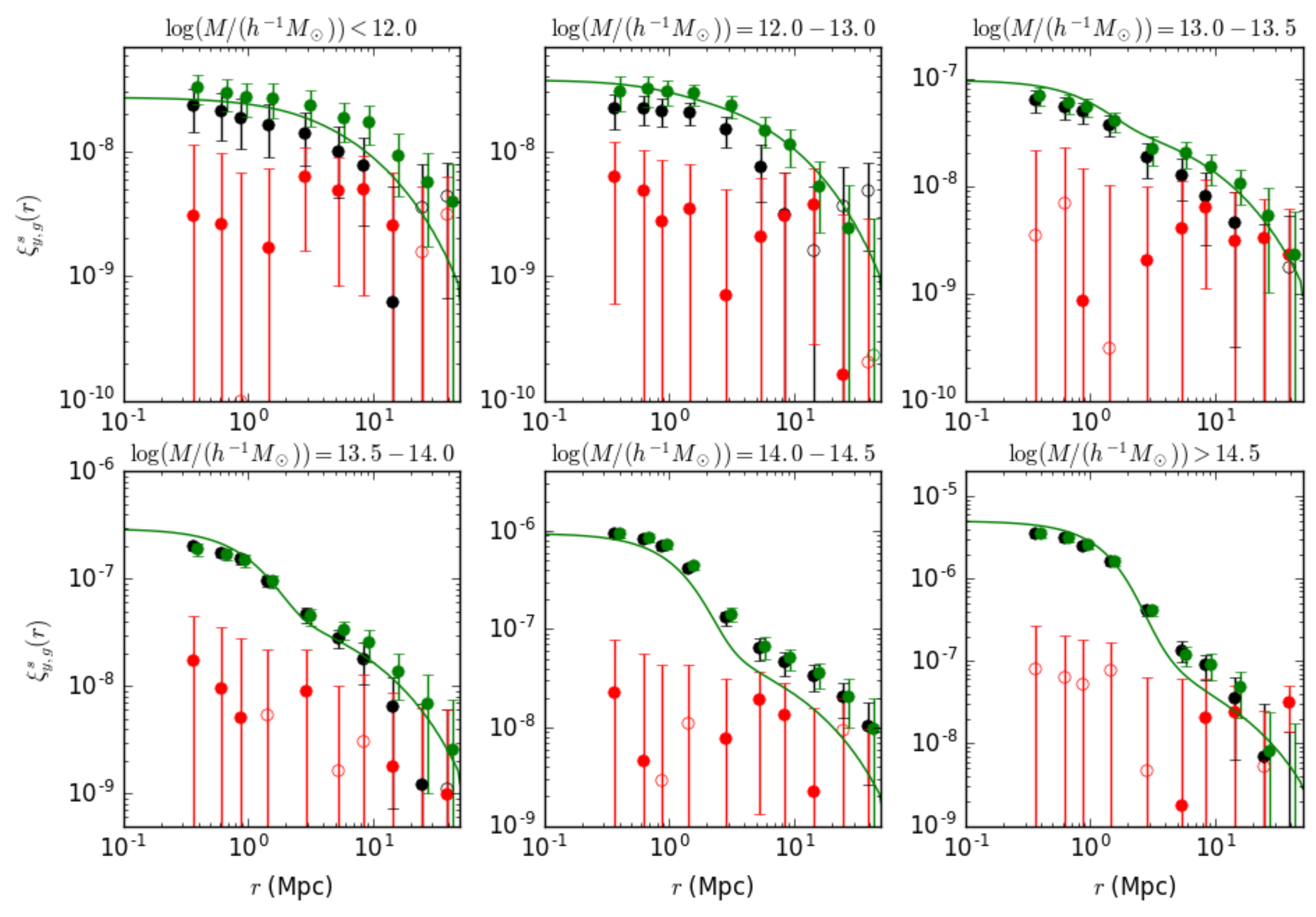}
\caption{Comparison of the SZ-group correlation function estimated from the Planck MILCA and NILC maps. The black points are correlation functions estimated from the NILC map, while
the MILCA-based measurements are the green points. The red points show the results from using the NILC null map. The green curve is our fiducial model prediction, as described in the previous
section. The MILCA points have been slightly offset along the direction of the x-axis for visual clarity.
}
\label{fig:nilc-xi}
\end{center}
\end{figure}

Another test is to measure the correlation function using the Planck NILC map; as we described in \S \ref{sec:ymap}, NILC is an alternate approach for reconstructing the Compton-\y signal from the Planck data. Comparing the correlation functions estimated from each of MILCA and NILC allows us to test sensitivity to some of the detailed assumptions made in extracting the \y signal from the data.  The results of this comparison are shown in Fig.~\ref{fig:nilc-xi}. In the two highest mass bins, the measurements are reassuringly similar, but the results differ somewhat in the other four mass bins. 
From inspection, it appears that the results generally agree above some threshold value of $\xi^s_{y,g}$, but
that the measurements begin to differ when $\xi^s_{y,g}$ falls below approximately $\xi^s_{y,g} \lesssim 10^{-8}$ or so, irrespective of mass bin. Since the typical noise per Planck pixel in the \y maps is of order $\sim$ a few $\times 10^{-6}$, we should keep in mind that the excess Compton-$\y$ signal we seek to measure is a tiny fraction of the noise, especially at large radii and around small mass groups. This places stringent requirements on the accuracy of our two-point function estimator. Although there do appear to be systematic differences between the results from the two maps, properly accounting
for correlations between the measurement errors in neighboring radial bins, the systematic shifts are fairly small relative to our present statistical errors. For example, we compute the
chi-squared difference between the two data sets to be $\chi^2_\nu=1.0$ in the lowest mass bin. In this calculation, we adopt the MILCA covariance matrix to describe the errors; we consider this error matrix alone rather than adding the two covariance matrices since the maps are not independent of each other. 
The MILCA/NILC differences are smaller in the other mass bins, especially in the two uppermost mass bins. In other words, the systematic differences appear to be at most comparable to the statistical error bars. This may seem surprising given the data points in Fig.~\ref{fig:nilc-xi}, but a ``chi-by-eye''
inspection gives a mis-leading impression owing to the strong correlations between the various radial bins (Fig.~\ref{fig:pm-xi-cov}).

In effort to diagnose the origin of these systematic differences, we turn to the Planck ``First Minus Last'' NILC map and use this as a null test for the Planck NILC-group correlation function results. 
Quantitatively, we find $\chi^2_\nu = 0.83, 1.27, 0.60, 0.75, 0.57, 1.50$, in order of increasing mass bin. Although these values are mostly larger than in the case of the MILCA map, the null
hypothesis still provides a reasonable fit to the measurement, at least in most of the bins. However, we can investigate whether a constant offset provides a better fit to the NILC null-group correlation function measurement. For
contrast, we also allow a constant offset fit to our MILCA null-group correlation function measurement.  As quantified in Table \ref{tab:null-bias}, the MILCA results tend to fare
a bit better in this test, with several NILC mass bins showing small preferences for non-zero offsets at a little more than $1-\sigma$ significance. 

The comparison between the MILCA and NILC results suggests that systematic errors remain in one or both of our measurements;
the differences identified, however, impact only the low mass bins and are at most
comparable to our statistical errors. It would be helpful to better understand the discrepancies here, but in this work we instead confine our main analysis to the Planck MILCA map, 
since this performs better in our null tests.
One difference between the two maps is that the noise in the Planck NILC map shows more prominent large-scale spatial variations (see Fig. 5
of \citealt{Aghanim:2015eva}). These variations might lead to a bias in the correlation function measurements if our random catalogs are insufficiently large/accurate. As we discuss in the next
sub-section, another possible issue relates to foregrounds leaking into the two \y-maps (at different levels), but this appears to be a sub-dominant concern. 

\begin{table}
\begin{center}
 \begin{tabular}{ccc}
$\log_{10}(M/(h^{-1} M_\odot))$ & MILCA  & NILC \\
\hline
$<$ 12.0 &  $-11 \pm 9.4 \times 10^{-10}$ & $1.5 \pm 1.4 \times 10^{-9}$\\
12.0-13.0 & $4.7 \pm 7.4 \times 10^{-10}$ & $2.0 \pm 1.2 \times 10^{-9}$ \\
13.0-13.5 &  $2.3 \pm 1.3 \times 10^{-9}$ & $3.0 \pm 2.0 \times 10^{-9}$ \\
13.5-14.0 & $ -7.7 \pm 18  \times 10^{-10}$ & $6.7 \pm 28  \times 10^{-10}$\\
14.0-14.5 & $-2.7 \pm 3.7  \times 10^{-9}$ & $1.5 \pm 5.2  \times 10^{-9}$\\
$>$ 14.5 &  $6.1 \pm 8.2  \times 10^{-9}$ & $1.3 \pm  1.1 \times 10^{-8}$\\
\end{tabular}
\end{center}
\caption{The best fit constant values to the SZ null-group correlation functions. Note that the errors are slightly larger for NILC map compared to the MILCA map. 
} 
\label{tab:null-bias}
\end{table}

\subsection{Bias from Residual Foregrounds}
\label{sec:cib_corr}

The Planck \y-maps are constructed to minimize foreground contamination, but the maps inevitably contain residual foregrounds owing, for the most part,  to imperfectly cleaned dust emission from our
own galaxy and from leaked CIB radiation. Here we follow previous work by \citet{hill14} and use the Planck 857 GHz map as a tracer of dust emission, and as a test
for residual dust emission in the Planck \y-maps. Further, we asses the impact of the leaked foregrounds on our \y-group correlation function measurements.
Since the emission in the 857 GHz map is completely dominated by the CIB and the galaxy, the emission at this frequency may be written as: 
\beqa
T_{857} (\theta) = T_{CIB}(\theta) + T_{gal}(\theta).
\eeqa
Here $\theta$ denotes the angular position on the sky and $T_{CIB}$ and $T_{gal}$ describe the CIB and galactic contributions to the emission. 
We suppose that a spatially-fixed fraction of these emissions leaks into the Planck \y-map\footnote{Spatial variations in the leakage coefficients would lead to smaller, higher order, corrections to our measurement and so we are justified in ignoring these variations.}, $\hat{y}(\theta)$, so that:
\beqa
\hat{y}(\theta) = y(\theta) + \alpha_{CIB} T_{CIB}(\theta) + \alpha_{gal} T_{gal}(\theta).
\eeqa
The symbol $\hat{y}$ here refers to MILCA's estimated Compton-\y parameter, while $y$ refers to the true \y parameter.
Here $\alpha_{CIB}$ and $\alpha_{gal}$ are coefficients that, respectively, describe the leakage of CIB and galactic emission into the MILCA \y map. Note that, using temperature units for
the emission in the 857 GHz map and for the CIB and galactic emission, these coefficients have units of inverse temperature.
 Since the emission from our own galaxy should be uncorrelated with the group sample\footnote{A caveat here is that we ignore an anti-correlation that may arise if it is significantly harder
 to identify groups behind regions of high galactic dust.},  the cross-correlation between the estimated \y map, $\hat{y}(\theta)$, and the group catalog,
 $n_g(\theta^\prime)$, is:
 \begin{align}
 \xi^s_{y,g} (|r - r^\prime|)  \equiv & \avg{\hat{y}(r) n_g(r^\prime)}  \nonumber \\
 = & \avg{y(r) n_g(r^\prime)} + \alpha_{CIB} \avg{T_{CIB}(r) n_g(r^\prime)}. \nonumber \\
 \label{eq:yn}
 \end{align}
 Here we have converted between angular coordinates, $\theta$ and $\theta^\prime$, and co-moving coordinates ($r$ and $r^\prime$) using the co-moving angular diameter distance to
 the redshift of each group. The above equation shows that our estimate of the SZ-group cross-correlation may be biased by the second term on the right hand side; however, this bias appears
 only to the extent that the \y map contains leaked CIB that itself correlates with the group catalog.

Following \citet{hill14}, we work in harmonic space and first consider the angular cross power spectrum between the Planck \y map and the 857 GHz map. This allows us to quantify the leakage from the CIB and galactic emission. Specifically, the
cross power spectrum between the \y and 857 GHz maps has an expected value of:
\beqa
C_\ell^{\hat{y}-857} = C_\ell^{y-CIB} + \alpha_{CIB} C_\ell^{CIB} + \alpha_{gal} C_\ell^{gal}.
\label{eq:857n}
\eeqa
The cross power spectrum hence has three separate contributions. The first piece arises because the CIB and the SZ are, to some extent, correlated with each other: i.e., the {\em true} \y map is partly
correlated with the CIB emission in the 857 GHz map. This correlation is described by the first term on the right hand side of Eq. \ref{eq:857n} and arises because there is {\em some} overlap between the
redshift windows of the CIB and SZ emission and so these two signals partly trace the same large scale structure. Their may also be a one-halo contribution to the CIB-SZ cross correlation if
the CIB and SZ signals originate, in part, from the same halos.
The CIB-SZ correlation term would be present even if the Planck \y map did not contain any residual
foreground contamination.
The second contribution is proportional to the auto power spectrum of the CIB emission (at 857 GHz), and owes to leaked CIB emission in the $\hat{y}$ map.  Finally, galactic emission that leaks into
the $\hat{y}$ map gives an additional term in the cross spectrum. These latter two contributions are described, respectively, by the second and third terms on the right hand side of Eq. \ref{eq:857n}. As
discussed in \citet{hill14}, this procedure assumes that the CIB emission in the lower frequency maps -- that enter our \y estimate -- is well-correlated with the the CIB at 857 GHz. This should be a good approximation. We also assume that the SZ signal itself makes a negligibly small contribution to the 857 GHz map.

We then measure the cross spectrum above and use it, along with a measurement of the 857 GHz auto-power, to estimate the level of CIB leakage in the Planck \y map. To do this, we use published models 
to specify $C_\ell^{CIB}$ and $C_\ell^{y-CIB}$ at 857 GHz. From the $C_\ell^{CIB}$ model and the 857 GHz auto power spectrum, we can determine the
angular power spectrum of the galactic foreground, $C_\ell^{gal}$ and vary $\alpha_{CIB}$ to best match the measured cross spectrum, $C_\ell^{\hat{y}-CIB}$. Previous studies in the literature have
determined the form of $C_\ell^{y-CIB}$ and $C_\ell^{CIB}$ by matching to multi-frequency measurements of the CIB angular power spectra, and also to the observed number counts of dusty star-forming
galaxies, from a range of different instruments \citet{Addison12,Addison13,Planck_CIB14,Planck_SZ_CIB15}. We use the most recent best fit determinations of $C_\ell^{CIB}$ from \citet{Planck_CIB14}
and $C_\ell^{y-CIB}$ from \citet{Planck_SZ_CIB15}. To estimate the power spectra from the Planck maps, we use the publicly available PolSpice code \citep{szapudi01} and mask $80\%$ of the sky. 
In estimating the leakage terms, we fit to cross spectrum measurements between $\ell=100$ and $\ell=1600$, but found similar values using the alternate range of $\ell=300-1600$.

Quantitatively, our best fit determination of the CIB and galactic leakage coefficients from the MILCA map are $\alpha_{CIB} = 1.22\times10^{-7} K^{-1}$ and  $\alpha_{gal} = -1.55 \times 10^{-6} K^{-1}$,
respectively. We find, however, that the $C_\ell^{y-CIB}$ term tends to dominate over the leakage terms in the observed $\hat{y}-857$ cross power spectrum on most angular scales.  
Since the error bars on $C_\ell^{y-CIB}$
are still sizable our leakage coefficient estimates have large error bars: unfortunately, as we will see, the procedure here does not allow a very tight bound on the correlated leakage.  
Specifically, the most recent determination of $C_\ell^{y-CIB}$  from the Planck collaboration is $1.6  \pm 0.7$ times their fiducial model (show in Fig. 2 of  \citealt{Planck_SZ_CIB15}).
We marginalize over this allowed range to determine error bars on the leakage coefficients, finding:
$\alpha_{CIB} = 1.22 \times 10^{-7} \pm 1.43 \times 10^{-6} K^{-1}$ and $\alpha_{gal} = -1.55 \times 10^{-6} \pm 9.10 \times 10^{-7} K^{-1}$. The best fit leakage model has  $\chi^2_\nu=0.62$  per degree of freedom , and is therefore an acceptable fit.

After estimating $\alpha_{CIB}$, the final step here is to to measure the cross-correlation between the Planck 857 GHz map and the group catalog. Since the galactic emission should not correlate with the group catalog (modulo the caveat mentioned previously), the expected value of this correlation is:
\beqa
\avg{T_{857}(r) n_g(r^\prime)} = \avg{T_{CIB}(r) n_g(r^\prime)}.
\label{eq:857_group}
\eeqa
Here we have again used the co-moving distance to the redshift of each group to convert between angles and co-moving distance units. Note also that we have moved back into configuration space, as we ultimately measure the two-point correlation function rather than the power spectrum here. From this measurement, and the $\alpha_{CIB}$ determined above, we obtain an estimate of the contamination term in Eq. \ref{eq:yn}. 

\begin{figure}
\begin{center}
\includegraphics[width=\columnwidth]{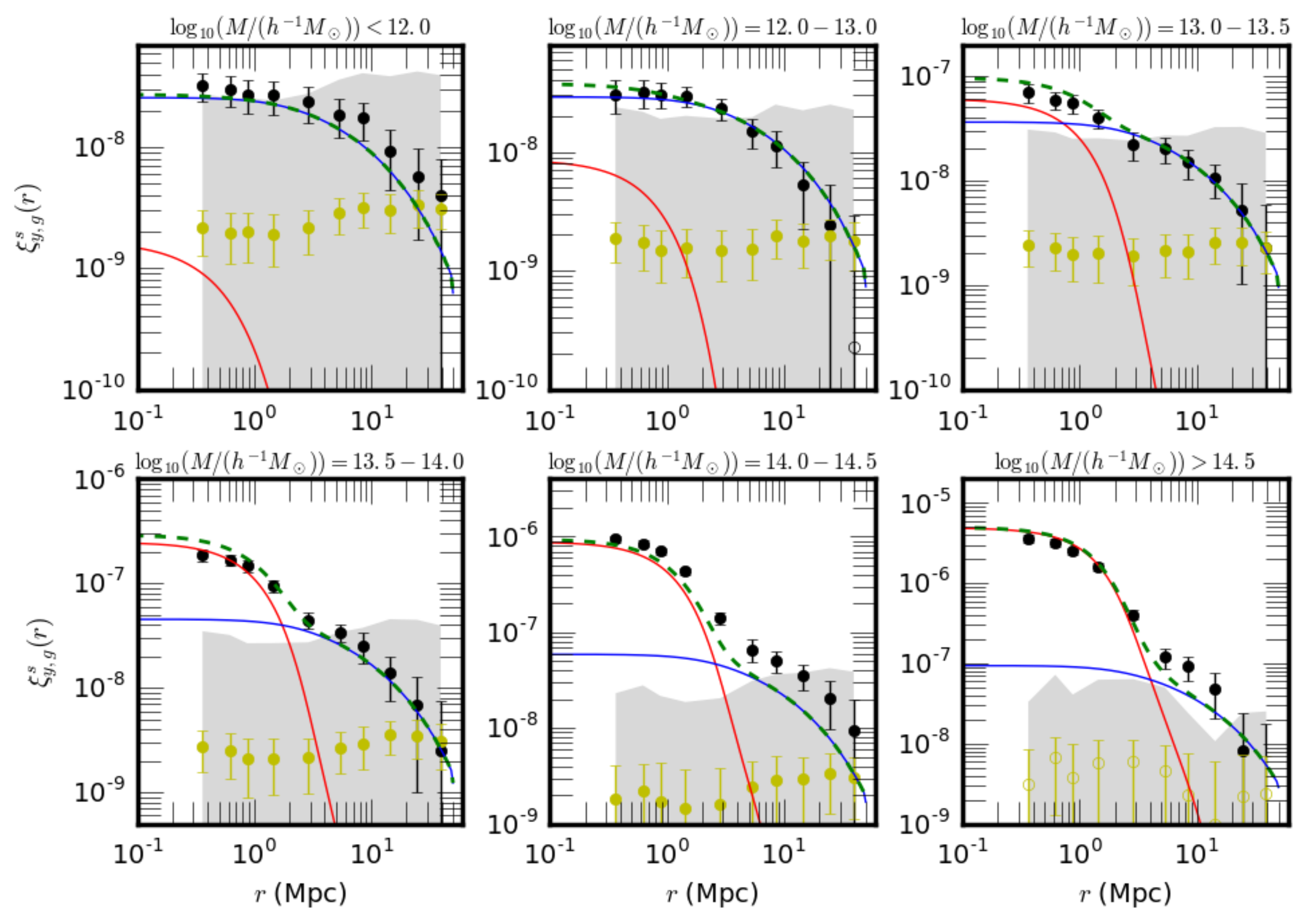}
\caption{Test for the impact of correlated CIB leakage on the measured SZ-group correlation function. The yellow points show our best fit estimate of $\alpha_{CIB} \avg{T_{857}(r)n_g(r^\prime)}$ from the MILCA map (see text), while the shaded region shows the allowed range in this quantity given the uncertainty in our determination of $\alpha_{CIB}$.  The error bars on the yellow points reflect uncertainties in our measurements of $\avg{T_{857}(r) n_g(r^\prime)}$, while the shaded band indicates the overall normalization
error.
The solid points correspond to positive values of $\alpha_{CIB}$, while the open points show negative values. The black points and colored lines are identical to those in Fig.~\ref{fig:pm-xi}. 
The fitting procedure allows a broad range of leakage coefficients, but the $\xi^s_{y,g}(r)$ measurements disfavor some of the larger values of $\alpha_{CIB}$ since these overproduce the measured SZ-group
correlation function. Nevertheless, CIB leakage is clearly an important systematic and refined estimates of $\alpha_{CIB}$ will be required to draw more definitive conclusions. 
}
\label{fig:systematic}
\end{center}
\end{figure}

The results of our leakage estimates are shown in Fig.~\ref{fig:systematic}. Considering first the best fit determination of $\alpha_{CIB}$ (shown as the yellow points), one can see that the preferred leakage term provides a highly
sub-dominant contribution to the overall $\xi^s_{y,g}(r)$ measurement for essentially all mass and radial bins. Interestingly, the best fit leakage term is a fairly flat function of scale, without a clear
one-halo to two-halo transition. This may be a consequence of the coarse Planck beam, mis-centering errors, and the contribution of satellite galaxies to the CIB emission.  Another intriguing feature
is that the sign of the CIB-group correlation seems to flip sign in the most massive bin. In any case, the best-fit CIB leakage estimate suggests that correlated leakage may not be a big contaminant to our
measurements; this is as one expects if the CIB is produced mostly by dusty star-forming galaxies at significantly higher redshift than our group catalog. However, the error bars on $\alpha_{CIB}$ are
large enough that our leakage estimates still allow strong contamination. On the other hand, this is at least partly disfavored by the measurement of $\xi_{y,g}^s(r)$ itself. If the leakage coefficient is too large --
even if still allowed by the $C_\ell^{\hat{y}-CIB}$ measurement -- this can overproduce $\xi_{y,g}^s(r)$, particularly in the low mass bins. We therefore suspect that the true CIB leakage coefficient must
not be much larger than our best fit value, although it may be smaller than the best fit. Nevertheless, given the large uncertainties in $\alpha_{CIB}$, our measurement may still be partly contaminated by CIB leakage.  Unfortunately, refined measurements of $\alpha_{CIB}$ are required to make more definitive statements. 

\section{Comparison with Previous Measurements}
\label{sec:comparison}

As mentioned in the introduction, our measurements were in part motivated by the Planck collaboration's analysis in \citet{Ade:2012nia} (see also \citealt{Greco15}). A natural question raised
by our  study is: could the Planck team's $Y_{500}$ measurements be biased from the two-halo contribution at low masses? Our results suggest that at masses below $10^{13.5} M_\odot$, the signal is indeed dominated by the two-halo term. 
The Planck analysis used an ``isolation criterion'' in effort to select only halos without nearby massive neighbors. 
Instead of attempting to find isolated groups, we measure the full group-SZ correlation and fit for the contribution from correlated neighbors. In this context, it is important to keep in mind that the massive systems produce a much larger SZ effect than the low mass ones, and so the one-halo contribution at low mass may be swamped
by the two-halo term, even when the excess probability of a massive neighbor is very slight.

As mentioned in the Introduction, the analyses of \citet{waerbeke14,ma15,hill14} are related to this work. In particular, \citet{waerbeke14} measures the SZ-weak lensing cross
correlation between Planck data and a CFHTLens mass map, while \citet{ma15} interpret this measurement using halo models. \citet{hill14} measure the SZ-CMB lensing cross correlation using Planck data, and explore the implications using a halo model approach. These analyses probe the SZ effect around large-scale structure at significantly higher redshifts than our measurement. 

\section{Future prospects}
\label{sec:future}
\begin{figure}
\begin{center}
\includegraphics[width=\columnwidth]{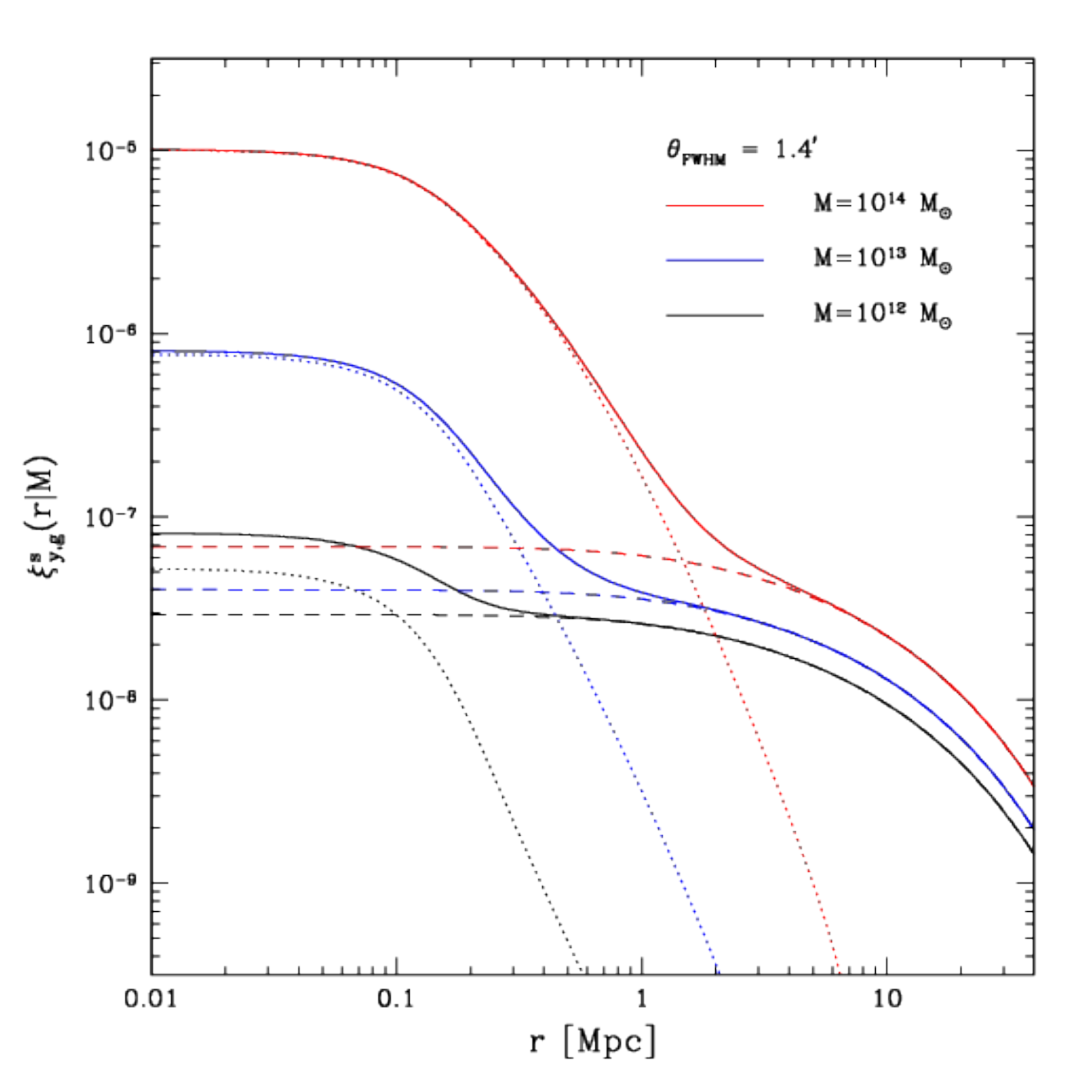}
\caption{The smoothed Compton-y halo cross correlation at ACT/SPT angular resolution. The red, blue, and black curves show the SZ-halo correlation function for halos of
mass $M=10^{14} M_\odot$, $M=10^{13} M_\odot$, and $M=10^{12} M_\odot$ (respectively) at $z=0.1$ for an angular resolution of $\theta_{\rm FWHM} = 1.4^\prime$. In each case, the
dashed curves show the two-halo term while the dotted lines give the one-halo contribution. }.
\label{fig:yprof_highres}
\end{center}
\end{figure}

In this section, we discuss the prospects for future  analyses that may help to overcome some of the
limitations of our current measurements. The first improvement we anticipate is in angular resolution: the dominance of
the two-halo term in our small-mass halo bins is in part a consequence of the coarse Planck beam.
Note that the virial radius of a $10^{13} M_\odot$ halo at $z=0.1$ subtends
an angle  $\Delta \theta_v = 4.8^\prime$, which is slightly less than half of the FWHM of the Planck beam (in the coarsest frequency channel). Fortunately, ACT and SPT can provide higher angular resolution CMB measurements. 

Fig.~\ref{fig:yprof_highres}  shows halo model predictions for the
 the Compton-\y map smoothed with $\theta_{\rm FWHM}=1.4^\prime$, comparable to the resolution of ACT and SPT, and a significant improvement over the Planck-beam value of $\theta_{\rm FWHM} = 10^\prime$.  We show predictions for halo
masses $M=10^{12} M_\odot$, $M=10^{13} M_\odot$, and $M=10^{14} M_\odot$, with the one and two-halo contributions shown separately.
With the higher resolution of ACT/SPT we can expect to detect the one-halo contribution for significantly smaller halo masses than Planck, possibly down to $M=10^{12} M_\odot$.  

There are, however, some caveats in the interpretation of upcoming high resolution CMB data. Future ACT and SPT cross-correlation analyses will need to rely on photometric data from the Dark Energy Survey (DES) and other surveys. 
An important issue, then, is
to understand how reliably group-finders may be applied to photometric data sets, especially in the low mass regime. 
Next, the reduced frequency coverage of ACT and SPT data in comparison to Planck will make it harder to separate the SZ effect from the CIB, and any other source of temperature anisotropy that may correlate with the group catalog. Fortunately, at least for low redshift, the CIB-group correlation is likely sub-dominant on small scales  (Fig.~\ref{fig:systematic}). In addition, we could extrapolate our multi-frequency
Planck measurement of the CIB-group correlation to small scales to quantify any contamination to measurements of the SZ-group correlation with ACT/SPT data although this may require
improved estimates of the leakage coefficient, $\alpha_{CIB}$.

\begin{figure}
\begin{center}
\includegraphics[width=\columnwidth]{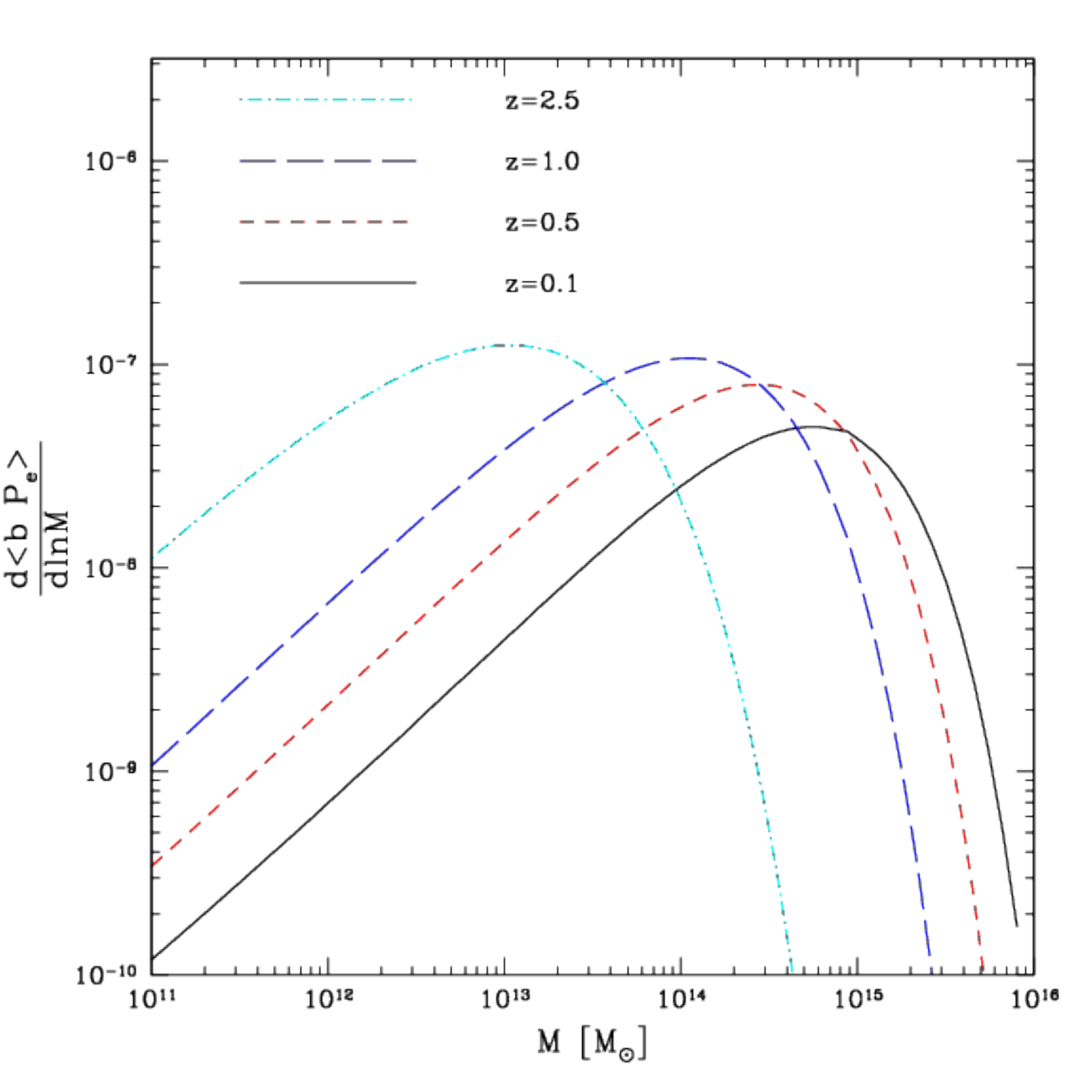}
\caption{The contribution to the average bias-weighted electron pressure from halos of different mass at a range of redshifts. Specifically, we plot the contribution to $\avg{b P_e}$ (see Eq. \ref{eq:bias_pressure}) per logarithmic interval in $M$ at each of $z=0.1, 0.5, 1.0$, and $2.5$ assuming the \citet{Battaglia:2011cq} pressure profile. As one moves to higher redshift, progressively
lower mass halos provide the dominant contribution to $\avg{b P_e}$ and the two-halo term probes systems that can not be detected directly in the SZ.}
\label{fig:bp_mass_cont}
\end{center}
\end{figure}

Another interesting direction is to measure the two-halo term across a wide range of redshifts by cross-correlating the   with additional tracers of large scale structure: quasars and various galaxy samples. This 
``SZ-tomography'' would be interesting for two reasons. First, this can help disentangle how much of the SZ background comes from hot gas at different redshifts. A second interesting -- yet connected -- point relates to {\em which systems} produce the dominant contribution to the two-halo term at various redshifts. 
Over the redshift range considered in this work, the model two-halo term is mostly contributed by hot gas in very
massive halos and indeed our results are consistent with the predictions of these models. Since we can probe the hot gas in these systems directly in the one-halo regime around massive halos,
the two-halo term does not here provide new information about hot gas in lower mass systems. The same is not, however, generally true at higher redshift: in this case, the two-halo term is sensitive
to hot gas in lower mass halos that can not be detected directly. Fig.~\ref{fig:bp_mass_cont} provides a quantitative illustration of this point: it shows that the peak contribution to the quantity $\avg{b P_e}$ --
which is probed in the two-halo regime (see Eq. \ref{eq:yprof_large_scale}) -- shifts to higher mass scales with decreasing
redshift, as structure grows hierarchically and the exponential cut-off in the mass function moves toward larger mass. For example, Fig. \ref{fig:bp_mass_cont} shows that the peak in $d\avg{b P_e}/d{\rm ln}M$ moves from $M_{\rm peak} = 5 \times 10^{14} M_\odot$ at $z=0.1$ to $M_{\rm peak} = 1 \times 10^{13} M_\odot$ at $z=2.5$. The high redshift case mentioned here may potentially be probed by cross-correlating Planck-based Compton-\y maps with BOSS quasar samples. This would provide an unprecedented opportunity to probe the hot gas in low mass halos at relatively early times. 
Although good statistics are required to measure the two-halo signal, since it is relatively weak, note that detailed group catalogs with halo mass estimates for each group are {\em not required}. Any
tracer of large-scale structure at the desired redshifts should suffice; the bias of the tracer -- which needs to be divided out to estimate $\avg{b P_e}$ from large scale measurements -- may be obtained from an auto-correlation measurement.

\section{Conclusions}

We have measured the SZ-group cross correlation function between Planck Compton-\y maps and the SDSS group catalog of \citet{yang07} and compared the measurement
with halo model predictions. Our results are mostly consistent with the simulated pressure profiles from \citet{Battaglia:2011cq} which incorporate AGN feedback. We explore the impact of plausible levels of mis-centering, mass scatter, and mass bias for the SDSS groups. The measurements do not yet provide a sharp test of these models, however, and we  summarize possible future improvements below. From the measured two-halo term, we determine
the bias-weighted electron pressure of the universe to be $\avg{b P_e} = 1.50 \pm 0.226 \times 10^{-7}$ keV cm$^{-3}$ at $z=0.1-0.2$. 

Our results suggest several possible directions for future work. 
First, galaxy-galaxy lensing measurements should help in determining the mis-centering, scatter, and bias
parameters for the group catalog, which otherwise limit our ability to determine the hot gas distribution around the groups. Second, high angular resolution data from ACT and SPT should allow us to determine
the gas pressure around lower mass systems. Third, cross-correlating with large-scale structure tracers at other redshifts should allow tomographic determinations of $\avg{b P_e}$ as a function
of redshift and thereby provide valuable information regarding the thermal energy content of the universe over cosmic time. Fourth, we found  systematic differences between our MILCA and NILC
results. While these are only comparable to the statistical errors for our present low mass measurements, a better understanding of the discrepancies  may be required for further progress. Along
these lines, improved CIB cleaning and characterization should help guard against correlated CIB leakage, which may lead to a bias especially at higher redshifts. 
Finally, it may be interesting to apply our methodology to measure the SZ-group two-point cross-correlation function for the locally brightest galaxy sample from \citep{Ade:2012nia}, and to use this to cross-check
their $Y_{500}$ measurements.

In terms of the modeling, it will be helpful to better calibrate our halo models using cosmological hydrodynamic simulations. Here it would be valuable to simulate the pressure profiles around lower mass halos than in most previous work, in order to further explore the expected signatures of non-gravitational heating at low mass. Next, the simulation based analyses can be provide superior models of the two-point correlation function in the
transition regime between the one and two-halo terms. Our current halo model makes simplifying assumptions, such as considering only linear biasing and neglecting halo exclusion effects, that should be improved on to best interpret future measurements, especially near this transition regime. We believe that extensions to the present work should lead to interesting constraints on the hot gas distribution around small mass systems and
on the thermal energy content of the universe as a function of cosmic time.

\label{sec:conclusion}

\section*{Acknowledgments} \label{sec:ThankYou} We thank Colin Hill for helpful comments on an early draft manuscript, and him, Nicholas Battaglia, Lindsey Bleem, Neal Dalal, Salman Habib, Katrin Heitmann, Mike Jarvis and Rishi Khatri for helpful discussions. We  thank Xiaohu Yang for providing his random group catalogs and Mike Jarvis for providing his TreeCorr code.  Argonne National Laboratory's work was supported under the U.S. Department of Energy contract DE-AC02-06CH11357. BJ is partially supported by the US Department of Energy grant DE-SC0007901. This work was performed at the Aspen Center for Physics, which is supported by National Science Foundation grant PHY-1066293


\bibliography{references}

\end{document}